\documentclass[fleqn,usenatbib]{mnras}

\usepackage{newtxtext,newtxmath}
\usepackage[T1]{fontenc}

\DeclareRobustCommand{\VAN}[3]{#2}
\let\VANthebibliography\thebibliography
\def\thebibliography{\DeclareRobustCommand{\VAN}[3]{##3}\VANthebibliography}

\usepackage{graphicx}	% Including figure files
\usepackage{amsmath}	% Advanced maths commands
\title[Testing afterglows of FRB 200428]{Testing afterglow models of FRB 200428 with early post-burst observations of SGR 1935+2154}

\author[A. J. Cooper et al.]{
A. J. Cooper$^{1,2}$ \thanks{E-mail: a.j.cooper@uva.nl},
A. Rowlinson$^{1,2}$,
R. A. M. J. Wijers$^{1}$,
C. Bassa$^{2}$,
K. Gourdji$^{3,7}$,
J. Hessels$^{1,2}$,\newauthor
A. J. van der Horst$^{4}$,
V. Kondratiev,$^{2}$
D. Michilli$^{5}$,
Z. Pleunis$^{6}$,
T. Shimwell$^{2,8}$,
S. ter Veen,$^{2}$
\\
$^{1}$ Anton Pannekoek Institute, University of Amsterdam, Postbus 94249, 1090 GE Amsterdam, The Netherlands\\
$^{2}$ ASTRON, the Netherlands Institute for Radio Astronomy, Oude Hoogeveensedijk 4, 7991 PD, Dwingeloo, The Netherlands\\
$^{3}$ Centre for Astrophysics and Supercomputing, Swinburne University of Technology, Hawthorn VIC 3122, Australia\\
$^{4}$ Department of Physics, George Washington University, 725 21st St NW, Washington, DC 20052, U.S.A. \\
$^{5}$ MIT Kavli Institute for Astrophysics and Space Research,  77 Massachusetts Ave, Massachusetts Institute of Technology, Cambridge, MA 02139, U.S.A. \\
$^{6}$ Dunlap Institute for Astronomy \& Astrophysics, University of Toronto, 50 St. George Street, Toronto, ON M5S 3H4, Canada  \\
$^{7}$ OzGrav: ARC Centre of Excellence for Gravitational Wave Discovery, Hawthorn, VIC 3122, Australia \\
$^{8}$ Leiden Observatory, Leiden University, P.O. Box 9513, 2300 RA Leiden, The Netherlands
}

\date{Accepted XXX. Received YYY; in original form ZZZ}

\pubyear{2022}

\begin{document}
\label{firstpage}
\pagerange{\pageref{firstpage}--\pageref{lastpage}}
\maketitle

\begin{abstract}
We present LOFAR imaging observations from the April/May 2020 active episode of magnetar SGR 1935+2154. We place the earliest radio limits on persistent emission following the low-luminosity fast radio burst FRB 200428 from the magnetar. We also perform an image-plane search for transient emission and find no radio flares during our observations. We examine post-FRB radio upper limits in the literature and find that all are consistent with the multi-wavelength afterglow predicted by the synchrotron maser shock model interpretation of FRB 200428. However, early optical observations appear to rule out the simple versions of the afterglow model with constant-density circumburst media. We show that these constraints may be mitigated by adapting the model for a wind-like environment, but only for a limited parameter range. In addition, we suggest that late-time non-thermal particle acceleration occurs within the afterglow model when the shock is no longer relativistic, which may prove vital for detecting afterglows from other Galactic FRBs. We also discuss future observing strategies for verifying either magnetospheric or maser shock FRB models via rapid radio observations of Galactic magnetars and nearby FRBs.
\end{abstract}


%%%%%%%%%%%%%%%%%%%%%%%%%%%%%%%%%%%%%%%%%%%%%%%%%%

%%%%%%%%%%%%%%%%% BODY OF PAPER %%%%%%%%%%%%%%%%%%

\section{Introduction}
Fast radio bursts (FRBs) are bright, millisecond duration flashes of coherent radio emission \citep{Lorimer2007,Thornton2013,Petroff2016,CHIME2021_catelog}. Over 20 FRBs have now been localized\footnote[1]{\url{http://frbhosts.org}} to their host Galaxies \citep{Chatterjee2017,Ravi2019,Bannister2019,Marcote2020,Nimmo2021,Heintz2020,Bhandari2020}, confirming that these bursts are (usually) of extragalactic origin, as was already implied by large dispersion measures. Two varieties of FRB source have been identified, categorized by whether the source is observed to repeat or not. Bursts observed from these different categories of sources appear to have different properties \citep{Pleunis2021a}, which may be suggestive of different progenitors or emission mechanisms.
\par
Since the discovery of FRBs, many theoretical models have been presented to explain the bursts involving various astrophysical sources and radiation mechanisms \citep{Platts2019}. Many of these models focus on highly magnetized neutron stars as prime progenitor candidates for a few reasons. Firstly, the millisecond emission timescale implies compact emission regions $R \approx c t_{\rm FRB} \Gamma^2 \approx 10^6 \, t_{\rm FRB,-3} \, \Gamma^2_{0} \; {\rm cm}$, where $\Gamma$ is the bulk Lorentz factor of the emission region towards the observer and we have used the convenient notation $X_{\rm n} \equiv X/10^n$. In addition, the luminosity of FRBs ($\approx 8$ orders of magnitude larger than giant millisecond radio pulses in the Milky Way) might imply they are powered by the large magnetic energy reservoir in the magnetospheres of highly magnetized neutron stars known as magnetars. For decades, magnetars within our Galaxy have been observed to emit spontaneous X-ray bursts, which appear to follow a similar wait time distribution and luminosity function to FRBs (e.g. SGR 1900+14 \citealt{Gogus1999}; SGR 1806-20 \citealt{Gogus2000}; FRB 121102 \citealt{Gourdji2019}), as well as periods of activity and quiescence. Finally, other observed properties of FRBs can be explained within a magnetized neutron star progenitor model: large rotation measures \citep{Michilli2018}, downward drifting sub-bursts (e.g. \citealt{Hessels2019}), and polarisation angle swings across bursts \citep{Luo2020}. 
\par
The exact radiation mechanism within a magnetar progenitor framework is also not understood, with the two primary classes of models distinguished by where the FRB is emitted: close to the surface of the neutron star in magnetospheric models \citep{Katz2016,Kumar2017,Lu2018,KumarBosnjak2020,Wadiasingh2019,2020arXiv200505093L}, and far away from the surface in models which rely on maser emission in magnetized shocks \citep{Beloborodov2017,Metzger2019,Beloborodov2020}. Both classes of radiation model appear to explain the basic properties of FRBs.
\par
Magnetospheric models of FRBs require the coherent radiation of accelerated particles from close to the neutron star surface, through e.g. curvature emission \citep{Kumar2017}. It has been shown that many of the observed properties of FRBs discussed above can be reproduced within this model (e.g. \citealt{Lu2018,Wang2019,Cooper2021a,Yang2021,Wang2022}). Coherent curvature radiation requires spatial inhomogeneities in the particle distribution (`clumps' or `bunches'), which may be created during the particle acceleration phase through the two-stream instability \citep{ChengRuderman1977,Usov1987,Lu2018,2022MNRAS.tmp.2283K}. However, whether clumps form on the required timescales is less clear (see \citealt{Melrose2021} and references therein for critical discussion of curvature radiation as a pulsar mechanism), and it has further been suggested that the FRBs produced within the magnetosphere may fail to propagate to an observer \citep{Beloborodov2021}.
\par
The maser shock model appears to be a theoretically robust way in which coherent emission can be produced, as the maser has been shown to emit narrow-band radio emission in 1D \& 3D particle-in-cell (PIC) simulations of magnetized, relativistic shocks \citep{Plotnikov2019,Sironi2021}. Furthermore, such emission could be ubiquitous in nature wherever such shocks occur, and non-magnetar progenitors have been invoked to explain observed periodicity in repeating sources \citep{Sridhar2021}. Such ubiquity could account for the high observed volumetric rate of FRBs \citep{Ravi2019,Lu2019,Luo2020}. However, the specific properties of recently reported FRBs are difficult to explain in a maser shock scenario. In particular, the microsecond variability of FRB 20180916B \citep{Nimmo2021} and the 30ms separation between distinct sub-bursts in Galactic FRB 200428 \citep{chime2020_sgr} appear to be contradictory to the large length scales associated with the synchrotron maser emission region. 
\par
In magnetar models of FRBs, multi-wavelength or multi-messenger counterparts to emission provide invaluable observing opportunities with which to distinguish between theoretical models. Both magnetospheric and maser shock models predict a high-energy counterpart to FRB emission \citep{Metzger2019,Cooper2021a,Yang2021}, however the large distances to extragalactic FRBs mean predicted X-ray/gamma-ray fluxes are below the detection threshold of current instrumentation. The first version of the synchrotron maser shock model \citep{Lyubarsky2014}, in which relativistic magnetar hyperflares ($U_B \approx 10^{48} \, {\rm erg}$) interact with pulsar wind nebula, predicted a very high-energy TeV component to FRBs. In \cite{Metzger2019} this model was adapted to consider interaction with a wind-like environment, or a mildly relativistic baryonic shell emitted from a previous flare. Using these models, \cite{Metzger2019} predict a multi-wavelength afterglow due to thermal synchrotron emission, based on gamma-ray burst (GRB) afterglow models \citep{Sari1998}. Notably, as the shocks required for the maser are highly magnetized \& relativistic, magnetic field lines are compressed such that shocks are quasi-perpendicular. For this reason, shock accelerated non-thermal electrons are not necessarily expected \citep{Sironi2009}, in contrast to what is observed in other astrophysical shocks. Observationally, identifying a multi-wavelength afterglow could confirm the synchrotron maser model of FRBs, since magnetospheric models are not predicted to produce them. Persistent radio counterparts have been identified in two localized repeating FRB sources: FRB 121102 \citep{Chatterjee2017} and FRB 20190520B \citep{Niu2021}. The origin of the persistent radio emission is not definitively known, the spectra and lack of variability may imply the compact source is an AGN that resides relatively close to the FRB source. Follow-up of all other localized FRBs have yielded only upper limits to afterglow counterparts \citep{Bhandari2018}, although these are generally not constraining for the maser shock model. 
\par
In this work, we present Low Frequency Array (LOFAR; \citealt{vanHaarlem2013}) imaging results of SGR 1935+2154, and discuss more broadly the multi-wavelength afterglow predicted after FRB 200428 in the maser shock model. In Section \ref{sect:sgr1935} we discuss the behaviour of magnetar SGR 1935+2154 before and during the 2020 active phase. In Section \ref{sect:LOFAR}, we present the LOFAR imaging results, where we perform a search for bright radio bursts observable in the image-plane (see \citealt{bailes2021} for the time-domain search, as well as other radio observations) and provide limits for persistent low-frequency emission. In Section \ref{sec:discussion} we interpret the LOFAR results and suggest observing strategies for verification of FRB emission mechanisms. In Section \ref{sec:maser_shock}, we discuss the synchrotron maser shock model and its application to FRB 200428. In particular, we find that early time upper limits appear to rule out simple versions of the maser shock model as applied to the Galactic FRB. We further discuss extensions to the model including non-thermal radiation at late times, and make afterglow predictions for Galactic and nearby, extragalactic FRBs. We conclude our findings in Section \ref{sect:conclusion}. 
% We put our results into context for models of FRBs that predict long-term variable and persistent emission, and discuss future prospects for detecting FRB afterglows at low frequencies. }

\section{SGR 1935+2154}
\label{sect:sgr1935}
Galactic magnetar SGR 1935+2154 was discovered in July 2014 through a series of short X-ray bursts \citep{Cummings2014,Lien2014,Israel2016}, and past X-ray activity from 2011 was found in an archival search \citep{Campana2014}. Further observations in 2015 with \textit{CHANDRA} \& \textit{XMM-Newton} found many more X-ray bursts \citep{Israel2016}, as well as the first intermediate flares \citep{Kozlova2016}. This led to the first measurements of a spin period and spin-down rate of $P = 3.245$ seconds \citep{Israel2016} and $\dot{P} = 1.43 \times 10^{-11}$, relatively typical values for a Galactic magnetar. Using these measured values, \cite{Israel2016} derive the characteristic age and surface magnetic field of the magnetar as: $\tau_{\rm c} = \dfrac{P}{2 \dot{P}} \approx 3.6$ kyr and $B_s = \sqrt{\dfrac{3 c^3 I P \dot{P}}{8 \pi^2 R_{\rm NS}^6}} \approx 2.2 \times 10^{14}$ G. Interestingly, $\dot{P}$ shows a decreasing trend, with a measured negative $\ddot{P} = -3.5(7) \times 10^{-19} \: {\rm s^{-1}}$ observed. The source position was determined to be at ${\rm R.A.} = 19:34:55.5978$, ${\rm Dec.} = 21:53:47.7864$, with an accuracy of 0.6'' (90\% confidence, \citealt{Israel2016}; see also the discovery of an infrared counterpart by \citealt{Levan2018} at Galactic coordinates ${\rm b} = 293.73128$, ${\rm l} = 21.896608$). The magnetar was associated with a supernova remnant SNR G57.2+0.8 by \cite{Gaensler2014}, with a full multi-wavelength radio study of continuum and persistent emission presented in \cite{Kothes2018}. The authors find that the age of the SNR is approximately $41$ kyr, i.e. much older than the characteristic age of the magnetar derived from its spin properties \citep{Israel2016}. The distance to SNR G57.2+0.8 was revised by \cite{Zhuo2020}, finding that the SNR and thus the magnetar is likely closer than the $10$ kpc previously assumed, at just $D = 6.6 \pm 0.7$ kpc.
% It was active again in late 2019 with detections of bursts from \textit{Swift} and \textit{Fermi} (Wood et al., GCN 25975; Ambrosi et al., GCN26153). 
\par
In late 2019 \citep{Wood2019,Ambrosi2019} and especially in early 2020 \citep{Palmer2020,Younes2020}, the magnetar entered a new phase of extreme activity. On the 27th April 2020 at 14:34:24, an extremely bright radio burst was emitted from SGR 1935+2154. The burst was observed by the Survey for Transient Astronomical Radio Emission 2 (\textit{STARE-2}; \citealt{Bochenek2020}) an all-sky radio telescope \citep{Bochenek2020a}. It was also detected in the side lobes of \textit{CHIME}, the Canadian Hydrogen Intensity Mapping Experiment telescope \citep{chime2020_sgr}. \textit{STARE-2} detected the burst with a $1.281-1.468$ GHz lower limit fluence of $1.5 \pm 0.3 \times 10^6$ Jy ms and a full-width half-maximum duration of $0.61$ ms. \textit{CHIME/FRB} observed two bursts separated by $28.9$ ms, with $400-800$ MHz durations of $0.585 \pm 0.014$ ms and $0.335 \pm 0.007 $ms respectively, and average fluences of $0.48 \pm^{0.48}_{0.24} \times 10^6$ Jy ms and $0.22 \pm^{0.22}_{0.11} \times 10^6$ Jy ms. The lower observed fluence could be attributed to a steep spectral rise with frequency such that the burst is brighter at the \textit{STARE-2} observing frequency. The implied luminosity was lower than extragalactic FRBs by $\approx 4$ orders of magnitude, but was brighter than any other coherent radio emission observed from within our Galaxy with the possible exception of the brightest 2 MJy nanoshot observed from the Crab pulsar \citep{Hankins2007}. On 30th April, a weak radio burst ($0.06$ Jy ms) was reported by \cite{Zhang2020}, and two further radio bursts were reported later by \cite{Kirsten2021} with fluences of $112$ Jy ms and $24$ Jy ms.
\par
FRB 200428 occurred during a very active period of high-energy bursts from SGR 1935+2154, and many hundreds of short X-ray bursts were observed by NICER, Fermi, Swift, AGILE, INTEGRAL, Insight-HXMT, Konus-Wind and other X-ray satellite observatories (Fig. \ref{fig:full_timeline}; for catalogues of X-ray bursts see \citealt{Younes2020,Cai2022,Li2022}). Remarkably, an X-ray burst temporally coincident with the FRB was observed by four X-ray instruments: Integral \citep{Mereghetti2020}, Insight-HXMT \citep{Li2021}, AGILE \citep{Tavani2021} and Konus-Wind \citep{Ridnaia2021}. The X-ray/radio energy ratio of the radio burst was $\dfrac{E_{\rm X}}{E_{\rm r}} \approx 10^{5}$. The radio-coincident X-ray burst had a harder spectrum than other bursts seen from SGR 1935+2154 in the same window of activity, extending to 250 keV \citep{Mereghetti2020,Li2021} despite a relatively typical total fluence. The stark contrast is clear when the X-ray burst is compared to the many Fermi-GBM and NICER observed bursts in \cite{Younes2021}. Within magnetospheric models of FRBs, this can plausibly be explained by a non-thermal counterpart observable only when an FRB is observed \citep{Cooper2021a}, in addition to a regular short magnetar burst; or due to rapid relaxation of perturbed magnetic field lines \citep{Yang2020}. In the synchrotron maser shock picture of FRBs, the hard X-ray burst represents the initial peak of the afterglow emission that is observable at lower frequencies at later times \citep{Margalit2020}.

\begin{figure}
\centering
\includegraphics[width=0.45\textwidth]{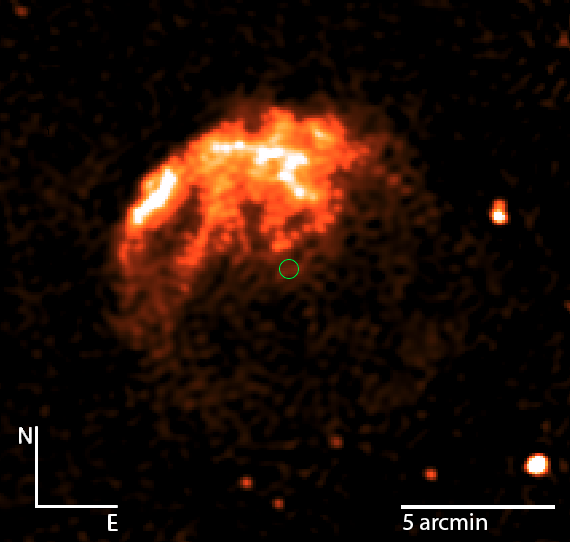}
\caption{LOFAR image of SGR 1935+2154 (denoted by the green circle) and the surrounding remnant SNR G57.2+0.8. This image makes use of the full datasets from all epochs.}
\label{fig:sgr1935_stack}
\end{figure}

\section{LOFAR observations}
\label{sect:LOFAR}

\begin{table}
\centering
\begin{tabular}{|c c c c|} 
\hline
Observation ID & Start Date & Start Time &  Calibrator ID \\
 & (UTC) & (UTC) & \\
\hline
L780243 & April 29 2020 & 03:08 & L780239 \\
L780251 & April 29 2020 & 05:20 & L780247 \\
\hline
L780651 & May 11 2020 & 01:49 &  L780655 \\
L780659 & May 11 2020 & 04:01 & L780655 \\
L780667 & May 11 2020 & 06:13 & L780671 \\
\hline
L797090 & October 21 2020 & 14:30 & L797092 \\
L797094 & October 21 2020 & 16:42 & L797096 \\
\hline
\end{tabular}
\caption[1]{The observations attained of SGR 1935+2154. The ID numbers correspond to the IDs for the preprocessed data in the LOFAR Long Term Archive\footnote[2]{\url{https://lta.lofar.eu}}. The duration of each target observation was 2 hours, separated by 10 minute calibrator observations of 3C295.}
\label{table:observations}
\end{table}

SGR 1935+2154 entered an active phase in April 2020. We obtained Directors Discretionary Time observations using LOFAR, comprising of 14 hours spread over 3 dates in 2020. The observation details are provided in Table \ref{table:observations}. Each observation was obtained using the LOFAR High Band Antennas (HBA) and the full Dutch array (24 core stations and 14 remote stations), covering a frequency range of 120--168 MHz and a central frequency of 144 MHz. The frequency range is divided into 244 sub-bands with bandwidths of 195.3 kHz. The recorded data have a time resolution of 1s, and a frequency resolution of 64 channels per sub-band. These data were pre-processed using the standard methods for LOFAR.

To calibrate these LOFAR data, we used the {\sc prefactor} pipeline (version 3.2)\footnote[3]{\url{https://github.com/lofar-astron/prefactor}} developed by the LOFAR Radio Observatory with the default parameters and version 3.1 of the LOFAR software. The calibrator source chosen for these observations was 3C48. As the field of SGR 1935+2154 is close to the A-Team source Cygnus A, we used the {\sc prefactor} demixing options to subtract the contribution of Cygnus A from the target data. The data were averaged to a time step of 8 seconds and a frequency resolution of 48.82 kHz (4 channels per subband).
\subsection{Epoch imaging}
All imaging of these data was conducted using {\sc WSClean} version 2.7.0\footnote[4]{\url{http://wsclean.sourceforge.net}} \citep{offringa2014}. The imaging settings used for these observations are outlined in Table \ref{table:wsclean}. Deep images of the field were created for the 3 observation days separately, as well as a fully combined image. Two versions of the deep images were created; one with the full dataset and the second with a minimum baseline length of 1000$\lambda$ to remove the extended emission from the supernova remnant. These images were used to search for faint persistent emission from the location of SGR 1935+2154. No persistent emission was detected at the location of SGR 1935+2154 and the limits obtained for persistent emission from SGR 1935+2154 are provided in Table \ref{table:limits}. We provide a constrained flux density measurement at the position of SGR 1935+2154 obtained using {\sc PySE} \citep{carbone2018} and assuming the restoring beam parameters. Additionally, we measure the rms noise in a 30$\times$30 pixel box centred on the position of SGR 1935+2154 and use the rms noise to calculate a 3$\sigma$ upper limit.

\begin{table}
\centering
\begin{tabular}{|c c c|} 
\hline
Paramter & Deep Images & Snapshot Images \\
\hline
Size (pixels) & 4096$\times$4096 & 2048$\times$2048 \\
Pixel Scale (arcsec) & 5 & 5 \\
-mgain & 10 & 10 \\
-j & 10 & 10 \\
Automatic masking of sources ($\sigma$) & 10 & 10 \\
Maximum baseline length ($\lambda$) & 8000 & 8000 \\
Minimum baseline length ($\lambda$) & 0/1000 & 700 \\
Number of frequency channels & 6 & 1 \\
Automatic threshold ($\sigma$) & 3 & 3 \\
Briggs weighting robustness & -0.5 & -0.5 \\
Number of iterations & 10000 & 10000 \\
Primary beam applied & True & False \\
Weighting rank filter & 3 & 3 \\
Clean border & 0 & 0 \\
Reorder visibilities & True & False \\
Fit restoring beam & True & True \\
Number of time intervals & 1 & 180 \\
\hline
\end{tabular}
\caption{The parameters used for imaging each observation set in {\sc WSClean}. Deep images are created using two different minimum baselines.}
\label{table:wsclean}
\end{table}

\begin{table}
\centering
\begin{tabular}{|c c c|} 
\hline
Observation Date & Flux density & 3$\sigma$ upper limit \\
 & (mJy) & (mJy/beam) \\
\hline
April 29 2020 & $2.2\pm5.8$ & 9.5 \\
May 11 2020 & $1.8\pm4.3$ & 6.8 \\
October 21 2020 & $0.7\pm4.6$ & 7.9 \\
\hline
All & $1.5\pm3.8$ & 6.0 \\
\hline
\end{tabular}
\caption{The flux density measurements at the location of SGR 1935+2154 for each observing epoch and a combined image. The flux density measurements were obtained via a constrained flux density measurement using {\sc PySE}. The 3$\sigma$ upper limits are obtained by measuring the rms noise in a region surrounding SGR 1935+2154.}
\label{table:limits}
\end{table}

\subsection{Snapshot Imaging}
To search for bright pulses from SGR 1935+2154, we also imaged the observations on shorter snapshot timescales. We integrated across the observed frequencies to create a single image. As SGR 1935+2154 has a high dispersion measure ($332.7 {\rm pc ~ cm^{-3}}$; \citealt{chime2020_sgr}), it takes time for a dispersed pulse to traverse the bandwidth of the HBA observation. This time is calculated using:

% \citep[332.7 pc cm $^{-3}$ ;][]{\citealt{chime2020}}, it takes time for a dispersed pulse to traverse the HBA observation. This time is calculated using:

\begin{eqnarray}
\tau_{\rm DM} = 8.3 \Delta\nu DM \nu^{-3} \mu{\rm s}, \label{eqn:dispersionDelay}
\end{eqnarray}

where $\tau_{\rm DM}$ is the dispersion delay in seconds, $\Delta\nu$ is the bandwidth in MHz, $\nu$ is the observing frequency in GHz, and $DM$ is the dispersion measure in pc cm$^{-3}$ \citep{taylor1993}. For these LOFAR observations, we find that the dispersion delay corresponds to 44 seconds. Therefore, we create snapshot images of 40 seconds in duration (as the data are integrated to 8 second intervals) and obtained 1081 images in total. We note that we could conduct de-dispersion imaging (e.g. \citealt{Anderson2021,Tian2022}) to obtain deeper constraints on very short duration dispersed pulses such as FRBs. However, beam-formed observations were also attained for these observations and very short duration dispersed pulses would be optimally detected in those observations \citep{bailes2021}. Our snapshot images targeted longer duration flares that may not have been detectable in the beam-formed searches. The snapshot images were created using the settings given in Table \ref{table:wsclean}. A minimum baseline length of 700$\lambda$ was used so that the extended emission from the supernova remnant was not visible in the snapshot images as this leads to a better constraint on emission from a point source at the location of SGR 1935+2154.

Following the method outlined in \cite{rowlinson2022}, we correct for any systematic flux density variations between the snapshot images produced by assuming that the majority of sources are not variable.  As SGR 1935+2154 is in the centre of the field, we consider sources out to a radius of 1 degree (covering the inner 50\% of the full width half maximum of the LOFAR HBA beam at 150 MHz\footnote[5]{\url{http://old.astron.nl/radio-observatory/astronomers/lofar-imaging-capabilities-sensitivity/lofar-imaging-capabilities/lofa}}). We extract all sources in the first 40 second snapshot image within that radius using {\sc PySE} with a detection threshold of 8$\sigma$ and confirm they are point sources by visual inspection (one source was rejected as it consisted of 2 blended point sources). These selection criteria resulted in 10 sources that were input into the LOFAR Transients Pipeline \citep[{\sc TraP};][]{swinbank2015} using the monitoring list capability. A flux density correction factor is then calculated and applied to each individual image\footnote[6]{using the script \url{https://github.com/transientskp/TraP_tools/blob/master/exampleScripts/correctSystematicFluxOffset.py}}. The average flux density correction is $\sim$6\%.

\begin{figure}
\centering
\includegraphics[width=0.49\textwidth]{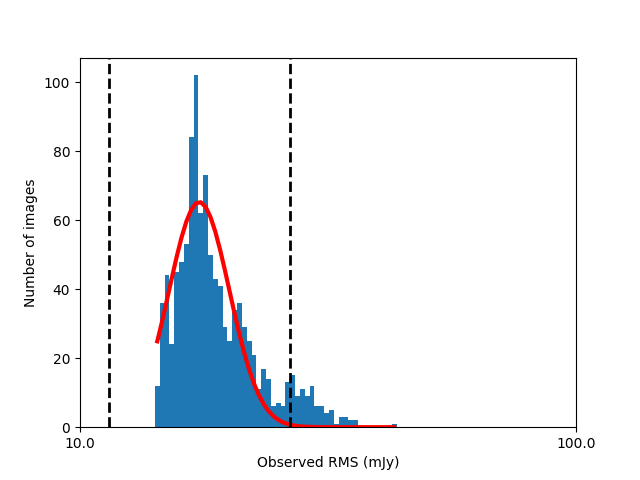}
\caption{Histogram of the rms values of the inner eighth of all the 40 second snapshot images used. The values are fit with a Gaussian distribution shown by the solid red curve. All images that are $>3 \sigma$ deviant from the average rms are rejected and the thresholds are shown with the black dashed lines.}
\label{fig:rms_histogram}
\end{figure}

Following the correction for systematic flux density variations, we conduct image quality control by measuring the rms noise in the inner eighth of each image and plot a histogram of these data in Figure \ref{fig:rms_histogram}. We fit the typical rms noise distribution for the images with a Gaussian distribution to give an average rms value of 17.4$^{+2.6}_{-2.3}$ mJy/beam. Any image with an rms noise that is more than 3$\sigma$ deviant from the average rms value is rejected. This analysis rejected 110 images, corresponding to 10\% of the sample.

\begin{figure}
\centering
\includegraphics[width=0.49\textwidth]{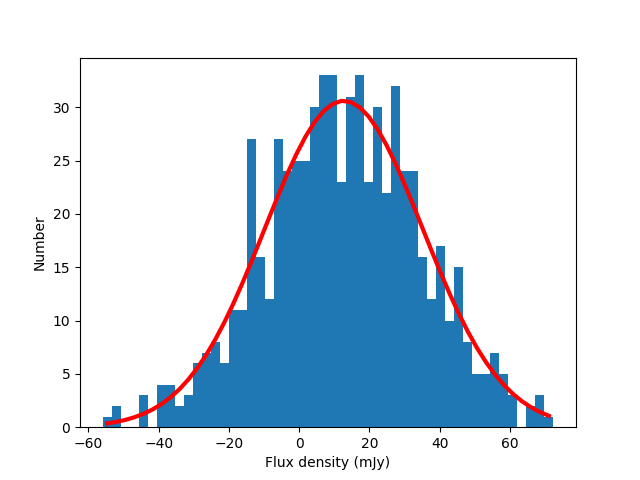}
\caption{A histogram of the flux density values for SGR 1935+2154 obtained by conducting a constrained fit at the location in each snapshot image.}
\label{fig:flux_histogram}
\end{figure}

To search for emission at the location of SGR 1935+2154, we processed the images using {\sc TraP} and monitor the position of SGR 1935+2154 to provide a flux measurement in each epoch. We used the standard {\sc TraP} settings apart from using a 4$\sigma$ detection threshold and setting the force beam parameter to True. As we are focusing on the location of SGR 1935+2154, which lies at the centre of the field, we use a small extraction radius of 120 pixels (corresponding to 10 arcmin). We detect no emission at the location of SGR 1935+2154. In Figure \ref{fig:flux_histogram}, we plot a histogram of the flux density values measured at the location of SGR 1935+2154 and fit the distribution with a Gaussian distribution. The fitted Gaussian distribution gives an average flux density measurement per epoch of 13$\pm$0.7 mJy with a standard deviation of 23 mJy, corresponding to a 3$\sigma$ upper limit for persistent point source emission of 82 mJy. This suggests the flux at the location of SGR 1935+2154 is significantly non-zero, which by inspection of Fig. \ref{fig:sgr1935_stack}, can be attributed to extended emission from the surrounding SNR.

\begin{figure*}
\centering
\includegraphics[width=1\textwidth]{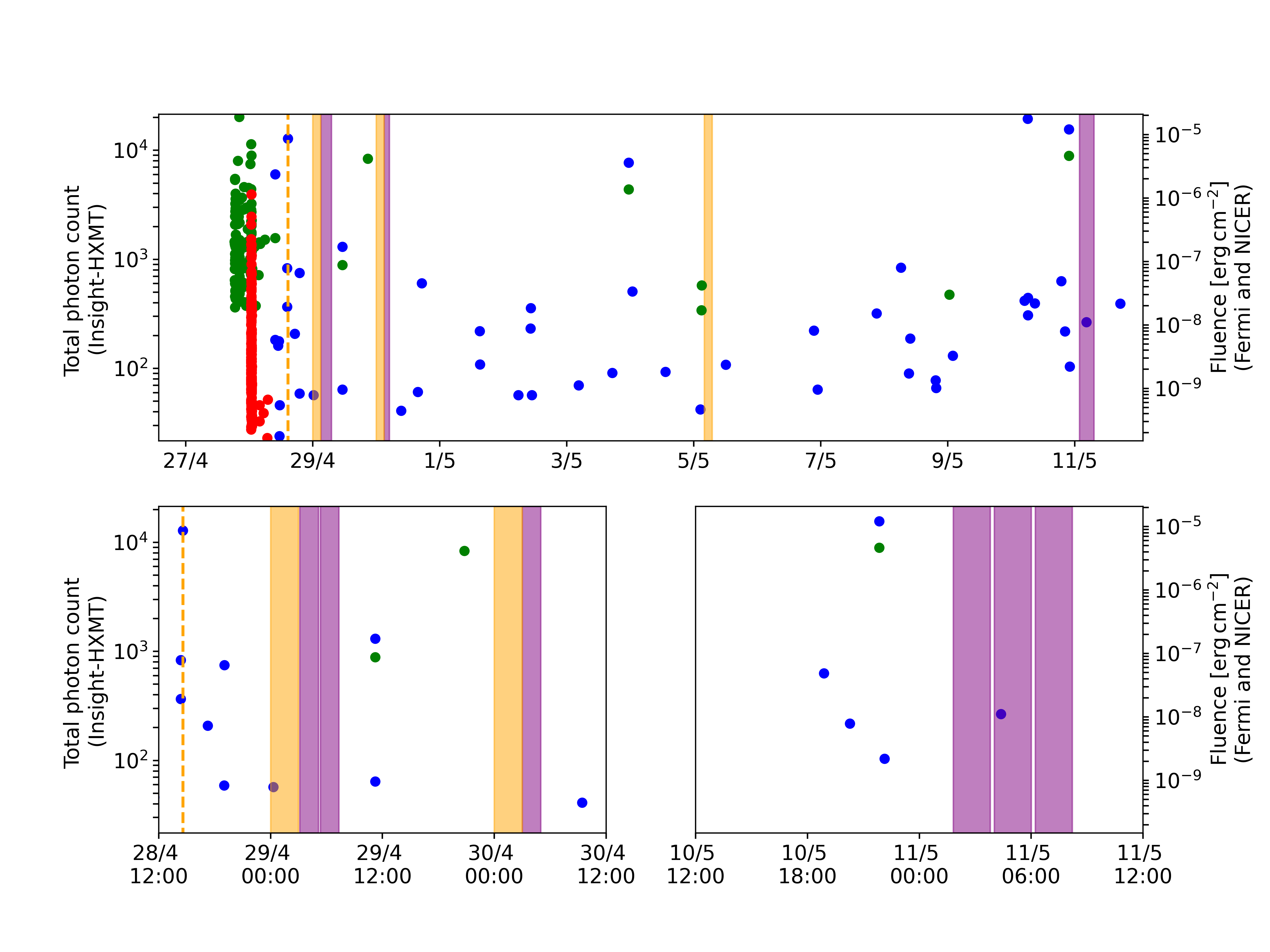}
\caption{We show the times of LOFAR observations (denoted by orange background hue) and AARTFAAC \citep{Prasad2016} observations (denoted by purple background hue) of SGR 1935+2154 during the April/May 2020 active episode. We note that observations on the 30th April were beamformed only, and therefore do not appear in our imaging analysis Section \ref{sect:LOFAR}. The time of the FRB as observed by CHIME \& STARE-2 is denoted by the vertical dashed orange line. We also show counts/fluence and times of X-ray bursts as observed by Insight-HXMT (blue dots; \citealt{Cai2022}), NICER (red dots; \citealt{Younes2020}) and Fermi-GBM (green dots; \citealt{Lin2020a}). Insight-HXMT data points refer to the total number of photon counts in the burst across all detectors (left y-axis), whereas NICER and Fermi-GBM data points refer to the total fluence of the burst in the 0.5-10 keV and 8-200 keV bands respectively. The second row of plots show timeline zoomed in on LOFAR observations, where 10 minute calibrator scans are visible, as are the two X-ray bursts for which we were on source. Interpretation of these observations is discussed in Section \ref{sec:discussion}.}
\label{fig:full_timeline}
\end{figure*}

%\begin{itemize}
%    \item TraP introduction and settings - usual settings and monitoring position of SGR 1935
%    \item TraP lightcurve analysis - similar to finding the GPs from a pulsar. Detections?
%    \item Flux limit in averaged images, single deep image (made by averaging data) and typical flux limit in 8 second images
%\end{itemize}

\section{Interpretation of LOFAR observations \& discussion}
\label{sec:discussion}

\subsection{Simultaneous radio \& X-ray limits on transient flares}
Low-frequency image-plane searches for radio transients have recently yielded detections of new types of coherent radio transient with long periods \citep{HurleyWalker2022,Caleb2022NatAstmp}. These new sources have inferred surface magnetic fields consistent with magnetars. We have performed a search for transient low-frequency emission in snapshot images on timescales of $40$ seconds (i.e. the dispersion delay across the LOFAR band of a burst from SGR 1935+2154) and find no transient emission at the location of the magnetar. Given the non-detections in the LOFAR time-domain search presented in \cite{bailes2021}, coupled with the fact that second-hour timescale transient radio emission has not been yet observed from Galactic magnetars, we did not necessarily expect a detection from this search.
\par
One of the primary motivators of radio observations of magnetars in active phases, other than detecting period-modulated radio pulses from magnetars (see Sect. \ref{sect:radio_loud_magnetars}), is to obtain constraints on the X-ray/radio luminosity ratio during X-ray magnetar bursts. In Fig. \ref{fig:full_timeline} we show the times of LOFAR and AARTFAAC observations, overplotted with high-energy bursts observed by various X-ray observatories. One high-energy burst observed by Insight-HXMT (burst number 56 in \citealt{Cai2022}) occurred during our LOFAR observations on 11/5/2022. The burst had a reported 1-250 keV flux of $2.29^{+0.18}_{-0.17} \times 10^{-7} \: {\rm erg \, s^{-1} \, cm^{-2}}$ \citep{Cai2022b}. The duration of this burst was $0.093$, $0.06$ and $0.076$ seconds in the high-energy (HE), medium-energy (ME) and low-energy (LE) detectors respectively, resulting in a X-ray fluence of approximately $2 \times 10^{-8}\: {\rm erg \, cm^{-2}}$ across 1-250 keV. The time-domain search for LOFAR bursts in this observation provided upper limits to the spectral fluence of a 1ms width radio burst of $0.075$ Jy ms \citep{bailes2021}. Assuming the bandwidth of a typical radio burst is $\delta \nu = 10^{9} \, {\rm Hz}$ (i.e. similar to the observed bandwidth of FRB 200428), we find a fluence limit of $7.5 \times 10^{-18} \: {\rm erg \, cm^{-2}}$, resulting in a minimum X-ray/radio fluence ratio of $\dfrac{\mathcal{F}_{\rm X}}{\mathcal{F}_{\rm r, 144 MHz}} > 3 \times 10^{9}$. This ratio is very constraining, as shock models of FRBs typically predict luminosity ratios of approximately $10^{5}$ \citep{Margalit2020}. 
\par
The Amsterdam-ASTRON Radio Transient Facility And Analysis Centre (AARTFAAC), is an all-sky radio transient monitor operating in parallel with LOFAR \citep{Prasad2016}, sensitive to bright transient and variable emission at low-frequencies. AARTFAAC was operating with 16 subbands spanning $60.15 - 64.84$ MHz during an X-ray burst observed from SGR 1935+2154 by Insight-HXMT (burst 19 in \citealt{Cai2022}). The dispersion delay from SGR 1935+2154 from the Insight-HXMT X-ray band to the AARTFAAC band is $\tau_{\rm X-ray} = 324$ seconds, and the in-band dispersion delay across $60.15 - 64.84$ MHz is $\tau_{\rm in-band} = 54$ seconds. No radio transient was detected at the time of the X-ray burst. We can derive a meaningful fluence sensitivity by comparing the weakest bursts observed in \cite{Kuiack2020} from PSR B0950+08 ($4 \times 10^{4}$ Jy ms), and rescaling by a factor proportional to the square root of the ratio of in-band dispersion delay from the different sources to account for a loss of sensitivity due to pulse smearing: $\bigg(\dfrac{\tau_{\rm PSR\, B0950+08}}{\tau_{\rm SGR\, 1935+2154}}\bigg)^{1/2} \approx 11$. We therefore find a conservative fluence sensitivity of approximately $10^{6}$ Jy ms for the dispersed burst from SGR 1935+2154, and we can constrain the fluence ratio of the X-ray burst observed by Insight-HXMT to $\dfrac{\mathcal{F}_{\rm X}}{\mathcal{F}_{\rm r}} \gtrsim 10^{2}$.
\par
Some magnetospheric models of FRBs predict that FRBs are always produced with short magnetar X-ray bursts, but are only beamed towards the observer in a small number of cases $\approx \Gamma^{-1}$ \citep{Lu2020}. Here, $\Gamma$ is the Lorentz factor of the emitting set of particles, and is on the order of $50-1000$ \citep{Kumar2017}. In the context of these models, coordinated radio and X-ray observations of active magnetars will either lead to the detection of another Galactic FRB, or build up a sample of radio-quiet magnetar bursts. If such a sample is significantly large (i.e. $N \gg \Gamma$) it can be used to statistically constrain $\Gamma$, and thus magnetospheric models of FRBs more broadly. Using single-dish 25m class telescopes, \cite{Kirsten2021} were able to place strong limits on radio counterparts to 59 high-energy bursts from SGR 1935+2154, effectively ruling out values of $\Gamma \ll 50$. Furthermore, \cite{Lin2020} found that 29 Fermi-GBM bursts were co-observed during one observing epoch with the FAST radio telescope, providing stringent upper limits on the radio/X-ray ratio of these bursts. The caveats with pursuing these kinds of simultaneous observation are two-fold. Firstly, we know that (repeating) FRBs generally have a narrow bandwidth (e.g. \citealt{CHIME2021_catelog}) and therefore observing at a one frequency does not necessarily preclude the emission of a radio burst at another. Second, magnetar X-ray bursts are stochastic and difficult to predict. Building such a sample will require either a pipeline from X-ray detections to radio observatories that enable rapid target-of-opportunity observations\footnote[7]{A possible observation strategy would be to use low-frequency radio telescopes and utilize the dispersion delay of radio emission to probe radio bursts associated with rapidly reported X-ray bursts. Alternatively, one may attempt to identify X-ray burst storms to trigger observations when more X-ray bursts are expected within the observation duration.}; or long-term observations of magnetars with underutilized radio telescopes. For the latter, we note that achieving a constraint of $\dfrac{L_{\rm X}}{L_{\rm r}} > 10^{5}$ for a typical Galactic magnetar bursts only requires fluence limits on the order of $100-1000$ Jy ms, and therefore does not require time on extremely sensitive radio instruments.

\subsection{Radio-quiet magnetars as FRB sources}
\label{sect:radio_loud_magnetars}
Pulsed, persistent emission is observed from a sub-population of magnetars which are radio-loud \citep{Camilo2006,Camilo2007,Kramer2007,Levin2010,Anderson2012,Lower2020,Esposito2020}. In \cite{Rea2012}, the authors describe the fundamental plane of radio magnetars, suggesting that in general radio-loud magnetars have quiescent $2-10$ keV X-ray luminosities below their rotational energy loss rate $L_{\rm rot} = 4 \pi^2 I P \dot{P}^{-3}$, where I is the neutron star moment of inertia. The conjecture that enhanced X-ray emission precludes radio emission (or vice versa) is supported by observations of PSR J1119-6127, a radio pulsar with a high magnetic field where radio pulses shut off when X-ray bursts are observed \citep{Archibald2017}. Similar X-ray and radio mode switches have also been observed in older (but non-recycled) pulsars with modest inferred magnetic fields, including PSR B0943+10 \citep{Hermsen2013} and PSR B0823+26 \citep{Hermsen2018}. 
\par
SGR 1935+2154 is thus far the only Galactic magnetar to have produced a bright radio burst, and the radio limits presented here and in \cite{bailes2021} suggest SGR 1935+2154 is not a radio-loud magnetar. This is consistent with the fundamental plane of \cite{Rea2012} given the observed persistent X-ray luminosity of SGR 1935+2154. Three individual radio bursts/pulses were reported after the FRB \citep{Zhang2020,Kirsten2021}, albeit at lower luminosities. The one-off radio bursts and pulses strongly suggest that high X-ray bursting activity does not disrupt a magnetar's ability to produce \textit{transient} radio bursts, and possibly that persistent pulsed radio emission precludes transient bursts. The association between radio-quiet magnetars and FRB-emitting magnetars is more intriguing when considered in the context of the recent discovery of highly magnetized neutron stars which exist beyond the canonical pulsar death lines (e.g. \citealt{Ruderman1975,ChenRuderman1993}) in $\dot{P}-P$ phase space (\citealt{Caleb2022NatAstmp}; see also \citealt{HurleyWalker2022}). The implication that a neutron star's inability to produce typical pulsar-like emission could be a prerequisite for a FRB production may hint that pulsed radio emission observed from (ultra) long period magnetars is powered by a different emission mechanism than pulsars, possibly more similar to the radiation mechanism that powers FRBs. Understanding the particle acceleration and coherent radiation mechanism operating in these long period sources may advance our understanding of how FRBs are generated.

\section{Maser shock model \& afterglow constraints}
\label{sec:maser_shock}
In the synchrotron maser shock model of FRBs, coherent radiation is generated by the gyration of particles in relativistic, magnetized shocks \citep{Lyubarsky2014,Beloborodov2017,Metzger2019,Beloborodov2020,Sironi2021}, as has been previously discussed in the literature before the discovery of FRBs \citep{Langdon1988,Hoshino1992,Gallant1992,UsovKatz2000}. The model suggests that flaring magnetars can produce relativistic ejecta, which provide conditions conducive to synchrotron maser emission upon interaction with external material in a surrounding nebula \citep{Lyubarsky2014}, or from a previous flare \citep{Metzger2019}. As the shock propagates relativistically, a multi-wavelength afterglow is expected \citep{Metzger2019}, with emission peaking at successively lower frequencies due to absorption and a decreasing bulk Lorentz factor $\Gamma$. 
\subsection{Afterglow model}
We present relevant dynamical \& radiation models, following the afterglow prescription detailed in \cite{Metzger2019} \& \cite{Margalit2020} in this subsection. We extend the model of \cite{Metzger2019} into the non-relativistic expansion phase relevant for late-time radio observations, remaining agnostic about the nature of the circumburst medium. Much of the early afterglow model is similar to models developed for GRB afterglows (e.g. \citealt{Meszaros1998,Sari1998}). The late-time dynamics of interest here after a deceleration time $t_{\rm dec}$ rely on the self-similar model of blast-waves described in \cite{BlandfordMcKee1976} and non-relativistic Sedov-Taylor model of an expanding blast-wave \citep{Sedov1959,Taylor1950}. Using these models we can make detailed predictions of the evolution of the shock and multi-wavelength afterglow.

\subsubsection{Deceleration phase}
As shown in \cite{Metzger2019}, the shock reaches the deceleration radius $r_{\rm dec}$ in a time $t_{\rm dec}$, which is equal to the central engine activity time, i.e. $t_{\rm dec} \approx t_{\rm FRB} \approx 1$ ms, such that we use these subscripts interchangeably. All parameters denoted with these subscripts refer to the value at the deceleration time. All timescales of interest for emission below the X-ray band occur at $t \gg t_{\rm dec}$, therefore we are only interested in solutions to the dynamics and radiation in the deceleration and non-relativistic phases. The bulk Lorentz factor for an adiabatically evolving shock at times $t \gg t_{\rm FRB}$ is:
\begin{equation}
    \begin{split}
        \Gamma(r > r_{\rm FRB}) &= \bigg(\frac{E_{\rm flare} (17-4k)}{16 \pi m_p n_{\rm ext} r^3 c^2} \bigg)^{1/2} \\
        &=\Gamma_{\rm FRB} \bigg( \frac{r}{r_{\rm FRB}} \bigg)^{\frac{k-3}{2}}
        \label{eq:gamma_int}
    \end{split}
\end{equation}
Where $r$ and $n_{\rm ext}$ are the radius of the shock from the central engine and the external number density of the local environment that the shock ploughs into. $n_{\rm ext}$ depends on the type of environment considered and in general has the form $n_{\rm ext} \propto r^{-k}$ where $0 \leq k < 3$. Wind-like medium ($k=2$) and constant medium ($k=0$) are commonly used, and either may be applicable to the environment surrounding a magnetar \citep{Metzger2019}. Assuming an external density at $n_{\rm ext}(r_{\rm FRB})$ at $r = r_{\rm FRB}$:
\begin{equation}
    n_{\rm ext}(r) = n_{\rm ext}(r_{\rm FRB}) \bigg( \frac{r}{r_{\rm FRB}} \bigg)^{-k}
% \end{equation}
% The total luminosity of the shock as seen by the observer is proportional to the mass swept up by the shock: 
% \begin{equation}
%     L_{\rm sh} = 4 \pi r^2 n_{\rm ext} \Gamma^4 m_p c^3
\end{equation}
In the observer frame, the distance of the adiabatic shock front from the central engine as a function of time is:
\begin{equation}
\begin{split}
     r(t > t_{\rm FRB}) &= 2 c \Gamma^2 t = 2 c t \Gamma_{\rm FRB}^2 \bigg(\frac{r}{r_{\rm FRB}} \bigg)^{k-3} \\
     &\propto t^{\frac{1}{4-k}}
     \label{eq:r_t}
\end{split}
\end{equation}
% Rearranging, we find that:
% \begin{equation}
% \begin{split}
%     %  r(t)^{4-k} &= 2 c t \Gamma(r_{\rm FRB})^2 r_{\rm FRB}^{3-k} \\
%      r(t) &=  \big(2 c t \Gamma(r_{\rm FRB})^2 \big)^{\frac{1}{4-k}} r_{\rm FRB}^{\frac{3-k}{4-k}}\\
%      \label{eq:r_t}
% \end{split}
% \end{equation}
Therefore, by Eqs. \ref{eq:gamma_int} \& \ref{eq:r_t} the shock Lorentz factor varies as a function of observer time as:
\begin{equation}
    \begin{split}
        \Gamma(r > r_{\rm FRB}) &= \Gamma_{\rm FRB} \bigg( \frac{1}{r_{\rm FRB}} \bigg)^{\frac{k-3}{2}} \bigg( \big( 2 c t \Gamma_{\rm FRB}^2 \big)^{\frac{1}{4-k}} r_{\rm FRB}^{\frac{3-k}{4-k}} \bigg)^{\frac{k-3}{2}} \\
        &\propto t^\frac{k-3}{8-2k}
        \label{eq:gamma_r_dec}
    \end{split}
\end{equation}
Given this, \cite{Metzger2019} compute the afterglow emission during this deceleration phase. Electrons in the shock are assumed to be in energetic equipartition with the ions such that their mean thermal Lorentz factor is: $\gamma_{\rm therm} = \dfrac{m_p}{2 m_e} \Gamma$. Particles are assumed not to undergo non-thermal acceleration by the shock (see Sect. \ref{sect:shock}). The magnetic field is parameterized in terms of the magnetization parameter $\sigma$ as a fraction of the thermal energy density of the shock such that:
\begin{equation}
    B(t > t_{\rm FRB}) = \sqrt{64 \pi \sigma \Gamma^2 m_p c^2 n_{\rm ext}} \propto t^{\frac{-3}{8-2k}},
    \label{eq:magnetization}
\end{equation}
where we assume throughout $0.1 \leq \sigma \leq 1$, as required for coherent maser emission. As in GRB afterglows, the critical synchrotron frequency $\nu_{\rm sync}$ and the cooling frequency $\nu_{\rm cool}$ \citep{Sari1998} in the observer frame is:
\begin{equation}
    \begin{split}
        \nu_{\rm sync} &= \frac{q B \gamma_{\rm therm}^2 \Gamma}{2 \pi m_e c} \propto t^\frac{3k-12}{8-2k}  \\
        \nu_{\rm cool} &= \frac{q B}{m_e c} \gamma_{\rm c, cool}^2 \Gamma \propto t^{\frac{3k-10}{8-2k}}
    \end{split}
    \label{eq:nu_sync_dec}
\end{equation}
% \begin{equation}
%     \begin{split}
%         \nu_{\rm cool} = \frac{18 \pi m_e c q B \Gamma}{\sigma_T^2 \Gamma^2 B^2 t^2}
%     \end{split}
% \end{equation}
% Therefore the observed luminosity for different regimes is:
% \begin{equation}
% L_{\nu}=
% \begin{cases}
% \frac{L_{\rm sh}}{2 \nu} \big( \frac{\nu}{\nu_{\rm cool}} \big)^{4/3} \big( \frac{\nu_{\rm cool}}{\nu_{\rm sync}} \big)^{1/2} \: \: \: \: \: \: \: \:\: \: \: \: \: \:  \: \: \: \: \: \: \:  \: \: \: \: \: \: \:  \: \: \: \: \:   \nu < \nu_{\rm cool},\nu_{\rm sync}\\
% \frac{L_{\rm sh}}{2 \nu} \big( \frac{\nu}{\nu_{\rm sync}} \big)^{1/2} \: \: \: \: \: \: \: \: \: \: \: \: \: \: \: \: \: \: \: \: \: \: \: \: \: \: \: \: \: \: \:  \: \: \: \: \: \: \:  \: \: \: \: \: \: \:  \: \: \: \: \: \nu_{\rm cool} < \nu < \nu_{\rm sync}\\
% \frac{L_{\rm sh}}{2 \nu} \big( \frac{\nu}{\nu_{\rm cool}} \big)^{4/3} \big( \frac{\nu_{\rm cool}}{\nu_{\rm sync}} \big)^{1/2} \exp \big( -\frac{\nu}{\nu_{\rm sync}} \big)  \: \: \:  \: \: \:  \: \:  \nu_{\rm sync} < \nu
% \end{cases}
% \label{eq:L_nu_dec}
% \end{equation}
Where $\gamma_{\rm c, cool} = \dfrac{6 \pi m_e c}{\sigma_{\rm T} \Gamma B^2 t}$. Using the parameter values of \cite{Margalit2020}, we find that for the flare associated with FRB 200428 from SGR~1935+2154, $\nu_{\rm sync} < \nu_{\rm cool}$ after a time of just $t = 0.99 \; \sigma_{-1}^2$ seconds. This means that for even the earliest optical data discussed in Section \ref{sect:optical}, the slow cooling regime holds. We note that for larger magnetization values (i.e. $\sigma = 1$) the earliest time predictions within a minute of the burst (e.g. Figs. \ref{fig:afterglow_optical} \& \ref{fig:afterglow_optical_windlike}) will be affected.
\par
The spectral luminosity at the critical synchrotron frequency is:
\begin{equation}
    L_{\nu, \rm sync} \propto N_{\rm therm} \Gamma B,
\end{equation}
where $N_{\rm therm} \propto n_{\rm ext} R_{\rm sh}^3$ is the number of radiating thermal electrons.
% \par
% The spectral luminosity at the critical synchrotron frequency is:
% \begin{equation}
%     L_{\nu, \rm sync} \approx \frac{L_{\rm sh}}{2 \nu_{\rm sync}}
% \end{equation}
In the slow-cooling regime \citep{Sari1998,Margalit2020}, the spectral luminosity is:
\begin{equation}
L_{\nu}=
\begin{cases}
L_{\nu, \rm sync} \big( \frac{\nu}{\nu_{\rm sync}} \big)^{1/3} \: \: \: \: \: \: \: \:\: \: \: \: \: \:  \: \: \: \: \: \: \: \: \: \: \:    \nu < \nu_{\rm sync}\\
L_{\nu, \rm sync}  \exp \bigg( \big( -\frac{\nu}{\nu_{\rm sync}}\big)^{1/3}\bigg)  \: \: \:  \: \: \:  \: \:   \nu > \nu_{\rm sync},
\end{cases}
\label{eq:L_nu_dec}
\end{equation}
where the $1/3$ exponent in the second line reflects recent theoretical work by \cite{MargalitQuataert2021}. In practice, for FRB 200428 we compute the predicted afterglow by fitting the initial X-ray data point and using the temporal scaling relations developed for GRB afterglow models described below \citep{Meszaros1998,Sari1998,Metzger2019}. To approximate the absorption of lower frequency emission, we follow the method of \cite{Margalit2020} used in their Fig. 9, by ensuring the spectral luminosity is not larger than the expected synchrotron self-absorption value of:
\begin{equation}
  L_{\nu, \rm SSA} \approx \frac{8}{3} \pi^2 m_e R_{\rm sh}^2 \nu^2 \gamma_{\rm therm}  
  \label{eq:absorption}
\end{equation}

\subsubsection{Non-relativistic phase}
The low luminosity of FRB 200428 means it enters a non-relativistic expansion phase much earlier than a bright FRB \citep{Margalit2020}. We can model this phase by comparison to the dynamics of supernova remnants, i.e. Sedov-Taylor expansion \citep{Taylor1950,Sedov1959}. We can find the time at which the shock enters this phase by considering when $\Gamma \approx 1$. Inspection of Eqs. \ref{eq:r_t} \& \ref{eq:gamma_r_dec} tell us $\Gamma(t > t_{\rm FRB}) \propto t^{\frac{k-3}{8-2k}}$ such that the approximate time and radius at which the shock becomes non-relativistic as:
\begin{equation}
\begin{split}
        t_{\rm non-rel} &\approx t_{\rm FRB} \bigg( \frac{1}{\Gamma_{\rm FRB}} \bigg)^\frac{8-2k}{k-3} \\
        r_{\rm non-rel} &\approx r_{\rm FRB} \bigg(\frac{t_{\rm non-rel}}{t_{\rm FRB}}\bigg)^\frac{1}{4-k}
        \label{eq:t_nonrel}
\end{split}
\end{equation}
For FRB 200428, we find that $t_{\rm non-rel} \approx 45$ seconds. We can find the synchrotron frequency and spectral luminosity of the thermal electrons at the time the shock is non-relativistic by Eq. \ref{eq:nu_sync_dec}:
\begin{equation}
        \nu_{\rm sync, non-rel} = \nu_{\rm sync,dec} \bigg(\frac{t_{\rm non-rel}}{t_{\rm FRB}}\bigg)^\frac{3k-12}{8-2k} 
\end{equation} 
In the slow-cooling regime, $L_{\nu, \rm sync} \approx$ constant in the deceleration phase \citep{Sari1998} such that:
\begin{equation}
    L_{\nu,\rm sync, non-rel} = L_{\nu,\rm sync, dec}
\end{equation}
After defining these quantities at the non-relativistic transition time, we can make predictions for this phase. Using the well-known Sedov-Taylor expansion solution in a constant medium, the shock radius in this regime is: 
\begin{equation}
     r(r > r_{\rm non-rel}) = r_{\rm non-rel} \bigg(\frac{t}{t_{\rm non-rel}}\bigg)^{2/5}
\end{equation}
In the slow-cooling non relativistic regime, the critical thermal synchrotron frequency and synchrotron luminosity are:
\begin{equation}
\begin{split}
    \nu_{\rm sync}(t > t_{\rm non-rel}) &= \nu_{\rm sync,non-rel} \bigg(\frac{t}{t_{\rm non-rel}}\bigg)^{-3} \\
    L_{\nu,\rm sync}(t > t_{\rm non-rel}) &= L_{\nu,\rm sync, non-rel} \bigg(\frac{t}{t_{\rm non-rel}}\bigg)^\frac{3}{5} 
\end{split}
\end{equation}
The above description assumes the thermal particles are relativistic, and therefore is valid while $\gamma_{\rm therm} \gg 1$. For FRB 200428, in the constant medium case the thermal electrons become non-relativistic approximately 1 day post-burst due to the low initial Lorentz factor. By this time the synchrotron cut-off frequency has dropped below the observing frequency for radio wavelengths of interest (see Fig. \ref{fig:sgr_1935_afterglow_lightcurve+limits}). However, extrapolating afterglow lightcurves beyond this time will require additional consideration of the deep-Newtonian regime similarly to GRB afterglows (e.g. \citealt{2015MNRAS.454.1711B}). Finally, we note that similar to SNR evolution, the shock is expected to enter the radiative snow-plough phase at a threshold velocity $v_{\rm sh} \approx \sqrt{\dfrac{k T}{m_e}} \approx 10^7 \, {\rm cm \, s^{-1}}$. For the constant medium case, this occurs for the FRB 200428 at approximately 1000 days post-burst affecting the predicted shock dynamics and lightcurves at very late times, which are not relevant for the observed upper limits.
\par
Using the above, we can make lightcurve predictions through the deceleration and non-relativistic phases for a variety of FRB bursts, as in \cite{Margalit2020a}. Following the prescription of \cite{Margalit2020}, we use the observed properties of FRB 200428 \& and the coincident X-ray burst to normalize the initial values $r_{\rm FRB}$, $\Gamma_{\rm FRB}$, $t_{\rm FRB}$ given the requirements of the maser and the high-energy burst. Crucially, the peak of the X-ray afterglow at $t_{\rm FRB}$ is normalized to the fluence and peak frequency observed by Insight-HXMT ($\mathcal{F} \approx 7 \times 10^{-7} \, {\rm erg \, cm^{-2}}$; $\nu_{\rm peak} \approx 50$ keV; \citealt{Li2021}). These parameters are outlined for FRB 200428 from SGR 1935+2154 in \cite{Margalit2020} (Eqs. 10-13; used in Figs. \ref{fig:sgr_1935_afterglow_lightcurve+limits} \& \ref{fig:sgr_1935_afterglow_lightcurve+powerlaw}). 

\subsection{Radio afterglow}
\begin{figure}
\centering
\includegraphics[width=0.45\textwidth]{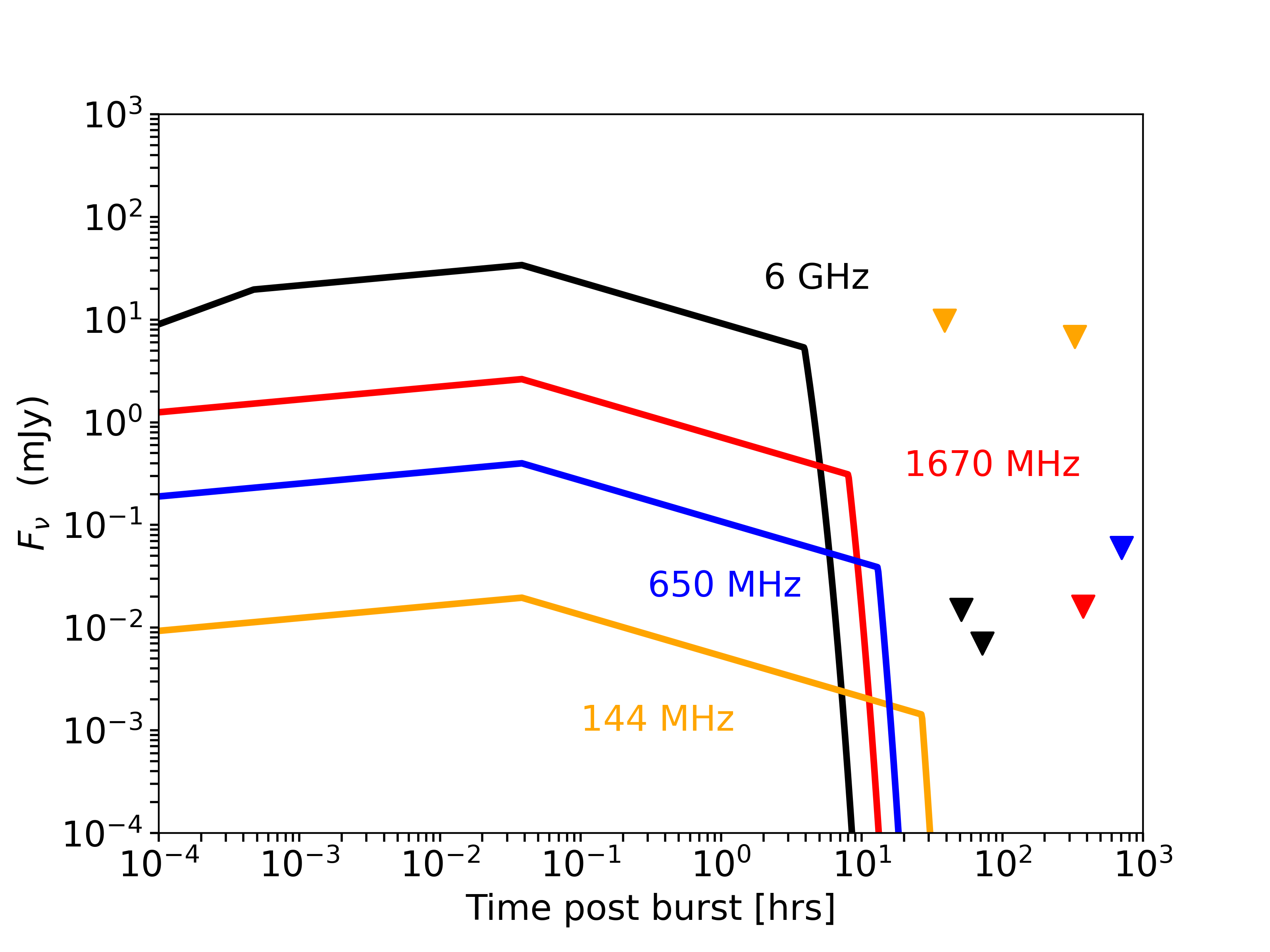}
\caption{Radio afterglow lightcurves following the FRB from SGR 1935+2154, following the prescription of \protect\cite{Margalit2020} in their Fig. 9 as discussed in the text. We show those frequencies with the most stringent and earliest radio observations \protect\citep{bailes2021}, including the LOFAR limits we present in this work. The lack of non-thermal particles leads to a steep decline in all lightcurves after just hours as the synchrotron frequency drops below the observing frequency.}
\label{fig:sgr_1935_afterglow_lightcurve+limits}
\end{figure}

In Fig. \ref{fig:sgr_1935_afterglow_lightcurve+limits}, we show predicted afterglow lightcurves for four frequencies for which radio limits were obtained within a short time after the initial burst \citep{bailes2021}, assuming a constant medium $(k=0)$. We find that that model-constraining observations could have been attained with the Karl G. Jansky Very Large Array (VLA; \citealt{Thompson1980}) \& European VLBI Network (EVN\footnote[8]{\url{https://www.evlbi.org/}}) at 6 and 1.67 GHz respectively if these instruments were on source before the synchrotron frequency of thermal electrons falls below the observing frequency. However, this would necessitate observations beginning just hours after the initial burst, which is incompatible with typical target-of-opportunity delay for large-scale radio facilities. In this case the obtained LOFAR limits presented in Section \ref{sect:LOFAR} are not very constraining, primarily attributable to the brightness of the extended emission from SNR G57.2+0.8 at $144$ MHz.

\subsection{Optical afterglow}
\label{sect:optical}
\begin{figure}
\centering
\includegraphics[width=0.45\textwidth]{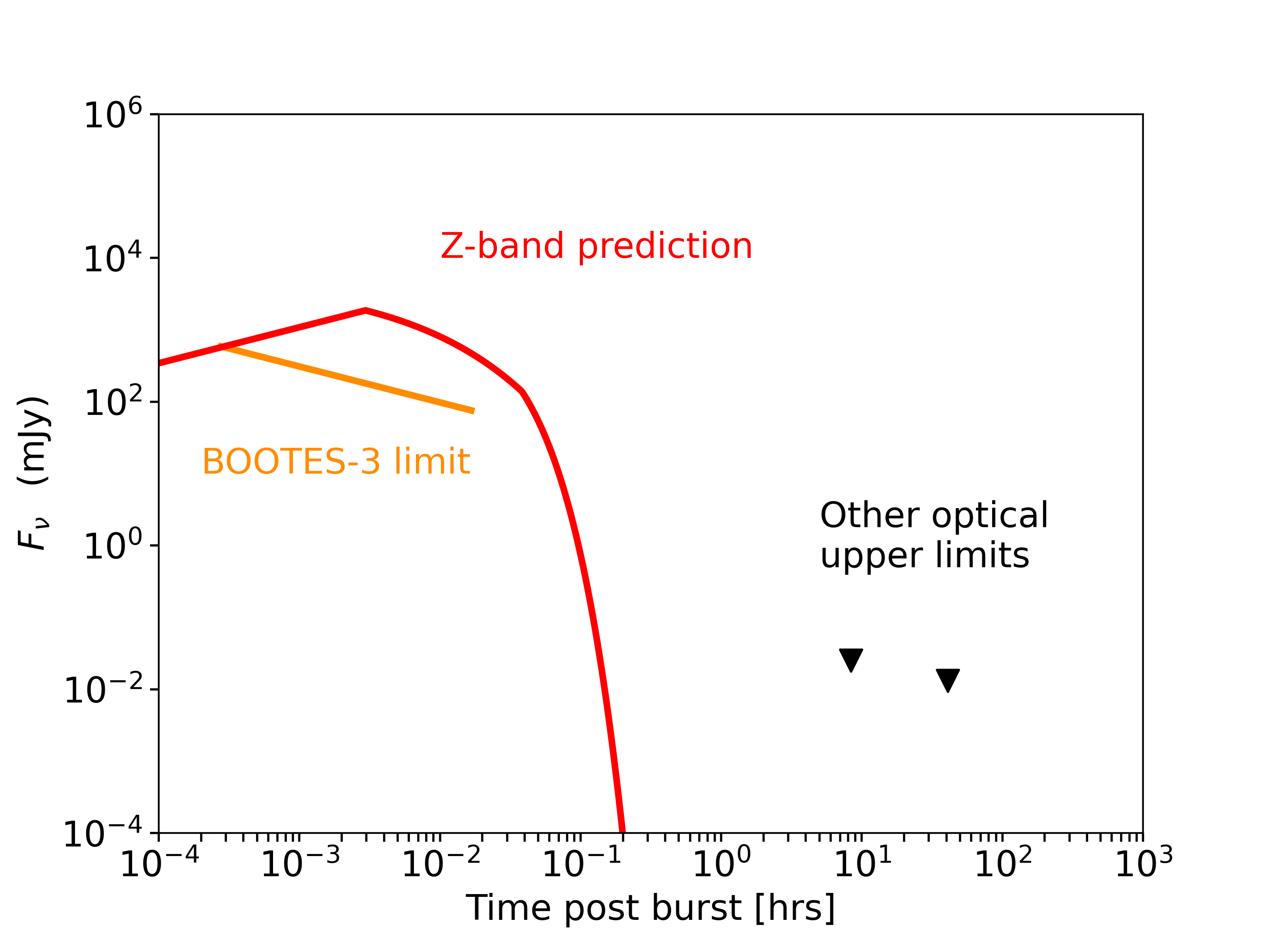}
\caption{We show in red the z-band lightcurve following the FRB from SGR 1935+2154, following the prescription of \protect\cite{Margalit2020}. In orange, the span of the observation taken by BOOTES-3 telescope (after 1 second on source), and in black we include additional optical upper limits by BOOTES-2 and LCOGT presented in \protect\cite{Lin2020}. The peak of the predicted flux is approximately one order of magnitude above the extinction-corrected limit set by the BOOTES-3 observation.}
\label{fig:afterglow_optical}
\end{figure}

In \cite{Lin2020}, the authors report on \textit{FAST} observations of SGR 1935+2154, as well as a multi-wavelength campaign spanning X-ray, optical and radio observations taken after FRB 200428. Of particular note is a minute-long z-band BOOTES-3 \citep{Castro-Tirado1999} observation that occurred simultaneously with the FRB 200428. The observation began on the 28th April 2020 at 14:34:24.03 (Extended Data Table 1; \citealt{Lin2020}), concluding a minute later and setting an upper limit of 17.9 mag. The authors revise this to just 11.7 mag when corrected for dust extinction of 6.2 mag, corresponding to a flux density of approximately 75 mJy. The peak of the X-ray burst coincident with FRB 200428 occurred after the optical observations began \citep{Mereghetti2020,Ridnaia2021}, meaning the optical limit is constraining for the very early time afterglow of the FRB. \cite{Lin2020} discuss the limit with respect to fast optical bursts (FOBs) predicted by \cite{Yang2019} to be produced coincident with FRBs. In Fig. \ref{fig:afterglow_optical}, we show z-band lightcurves using the above method (see also Fig. 9 in \citealt{Margalit2020}), and note that the BOOTES-3 upper limit significantly constrains the FRB afterglow. We assume that the flux limit scales as $F_{\rm lim} \propto (T_{\rm obs})^{-1/2}$ and that limits could be placed 1 second after the FRB and the start of the observations. We note that there are uncertainties with extrapolating the upper limit to the start time of the observation in this manner. Even without such extrapolation, the reported data strongly suggests that the predicted afterglow of the maser shock model in a uniform medium presented in \cite{Margalit2020} is not compatible with the observed upper limit.

\subsubsection{Wind-like case}
One way in which the maser shock model could circumvent the optical upper limit is to invoke a more complex, non-uniform environment. In this subsection, we consider the FRB 200428 afterglow shock propagating into a wind-like ($k=2$) medium at early times. To correct the initial values of $\Gamma_{\rm FRB}$ \& $r_{\rm FRB}$ we refer to the wind-like case in Section 2.2.3 of \cite{Metzger2019}. 
\begin{equation}
    \Gamma_{\rm FRB} = 2.8 \: E_{\rm flare}^{1/4} \, \dot{M_{19}}^{-1/4} \, \beta_{\rm W}^{1/2} \, \delta t^{1/2}
\end{equation}
\begin{equation}
    r_{\rm FRB} = 1.5 \times 10^{9} \: {\rm cm} \; E_{\rm flare}^{1/2} \, \dot{M_{19}}^{-1/2} \, \beta_{\rm W}^{1/2} \, \delta t^{1/2}
\end{equation}
Where $E_{\rm flare}$ is the total energy of the flare, $\dot{M}$ is the mass injection rate in units of grams per second, and $\beta_{\rm W} = v_{\rm w}/c$ is the time-averaged magnetar wind velocity divided by the speed of light. We note that the Lorentz factor required to explain the low-luminosity X-ray flare is relatively low, challenging the implicit assumption required for maser emission that the flare is initially ultra-relativistic. As a lower limit to the flare energy $E_{\rm flare}$, we adopt the X-ray luminosity of the coincident flare so that $E_{\rm flare} = 7 \times 10^{39}$ erg. We also consider a range of values for the unknown mass injection rate for this source: $\dot{M} = 10^{17-21}$ g s$^{-1}$. Values of $\dot{M} = 10^{19-21}$ g s$^{-1}$ have been shown to be consistent with the persistent radio nebula and rotation measure of FRBs from FRB 121102 \citep{MargalitMetzger2018}. However, we also consider values $\dot{M} < 10^{19}$ g s$^{-1}$ due to the lack of any persistent radio nebula at the location of SGR 1935+2154 in Section \ref{sect:LOFAR}, and the fact that this source does not appear to be a prolific FRB emitter. In the non-relativistic regime, we use the Sedov-Taylor expansion solution corrected for wind-like media such that $R \propto t^{2/3}$. 
\par
\begin{figure}
\centering
\includegraphics[width=0.45\textwidth]{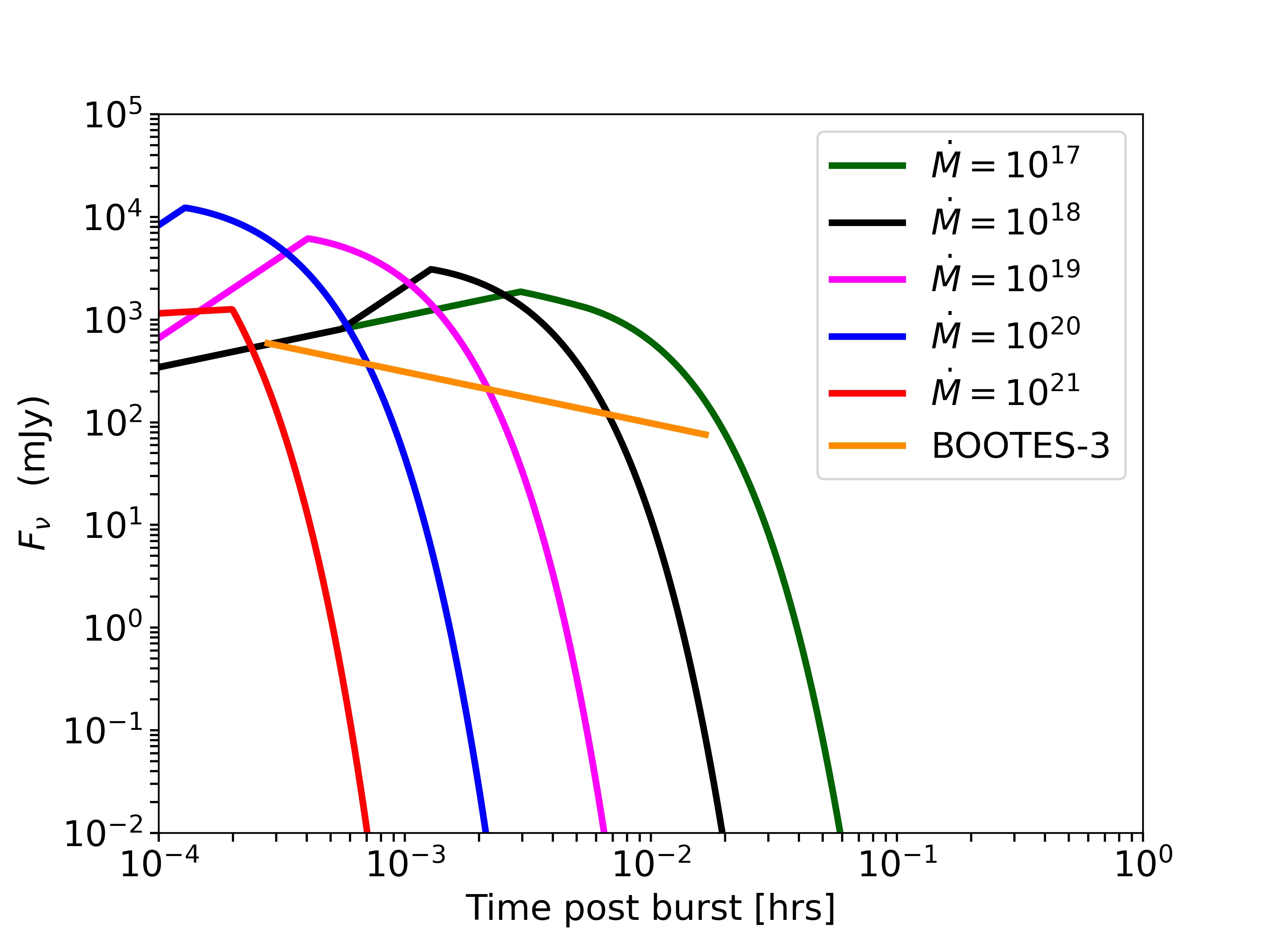}
\caption{Afterglow z-band lightcurves the wind-like environment case for various values of the mass injection rate $\dot{M}$. The flux sensitivity of the BOOTES-3 limit is assumed to scale as $T_{\rm obs}^{-1/2}$ and we show only limits from one second after the start of the observation. This appears to rule out the most values of the mass injection rate, however extrapolation of the observations to the time of the FRB (where they are reported to begin) may completely rule out the maser shock model of FRB 200428.}
\label{fig:afterglow_optical_windlike}
\end{figure}

In Fig. \ref{fig:afterglow_optical_windlike} we show the z-band predictions for the wind-like medium case, considering a range of values for $\dot{M}$. We also show the flux limits obtained by BOOTES-3, again assuming that the flux limit sensitivity scales as $T_{\rm obs}^{-1/2}$ from one second after the start of the observation. We find that the optical limits are consistent with scenarios in which $\dot{M}>10^{19}$ g s$^{-1}$, but caution that a more detailed analysis of the earliest second of the optical dataset may rule out all values of $\dot{M} < 10^{21}$ g s$^{-1}$. Finally, we note that $n \propto R^{-k}$ where $k > 2$ might be expected during the early-time deceleration phase at the time of the BOOTES-3 observation. This is because the charge density of the magnetar's magnetosphere (which has a radial dependence of $k \approx 3$) likely contributes to the local density at $r \approx 10^{9}$ cm. Estimated in terms of the Goldreich-Julian density \citep{Goldreich1969}, the contribution is:
\begin{equation}
   n_{\rm GJ}(r_{\rm FRB}) \approx \frac{2 B_s}{c q P} \bigg(\frac{R_{\rm NS}}{R} \bigg)^{-3} \approx 10^{4} \; {\rm cm^{-3}} \; B_{\rm s, 14}\, P_{0}^{-1}
   \label{eq:goldreich_julien_density_magnetar_a}
\end{equation}
This is comparable to the required value of $n_{\rm ext} = 4 \times 10^{4} \, {\rm cm^{-3}}$ in the maser shock model of FRB 2000428 in Eq. 12 of \cite{Margalit2020}, therefore considering variable values of $k$ as a function of $r$ may be required for a more accurate description of the afterglow.

\subsection{Prospects of radio afterglow detection for the next Galactic FRB}
As aforementioned, the lower limit to the radio luminosity of the Galactic FRB 200428 from SGR 1935+2154 was approximately 4-5 orders of magnitude less than typical FRB luminosities. In this subsection, we discuss strategies with which to detect an afterglow for a future Galactic FRB. 
\par
A bright FRB occurring within the Galaxy could be detected by all-sky radio telescopes such as STARE-2\footnote[9]{STARE-2 is now decommissioned, but a successor instrument to the project is planned: \cite{Connor2021}} or AARTFAAC \citep{Prasad2016}, or in the side-lobes of smaller field-of-view radio telescopes. Furthermore, the prompt X-ray component of the afterglow (or coincident magnetar X-ray burst) will be very bright and possibly detected by wide-field gamma-ray instruments such as \textit{Fermi-GBM}, or other X-ray instruments as in the case of the burst from SGR 1935+2154. 
\par
\begin{figure}
\centering
\includegraphics[width=0.45\textwidth]{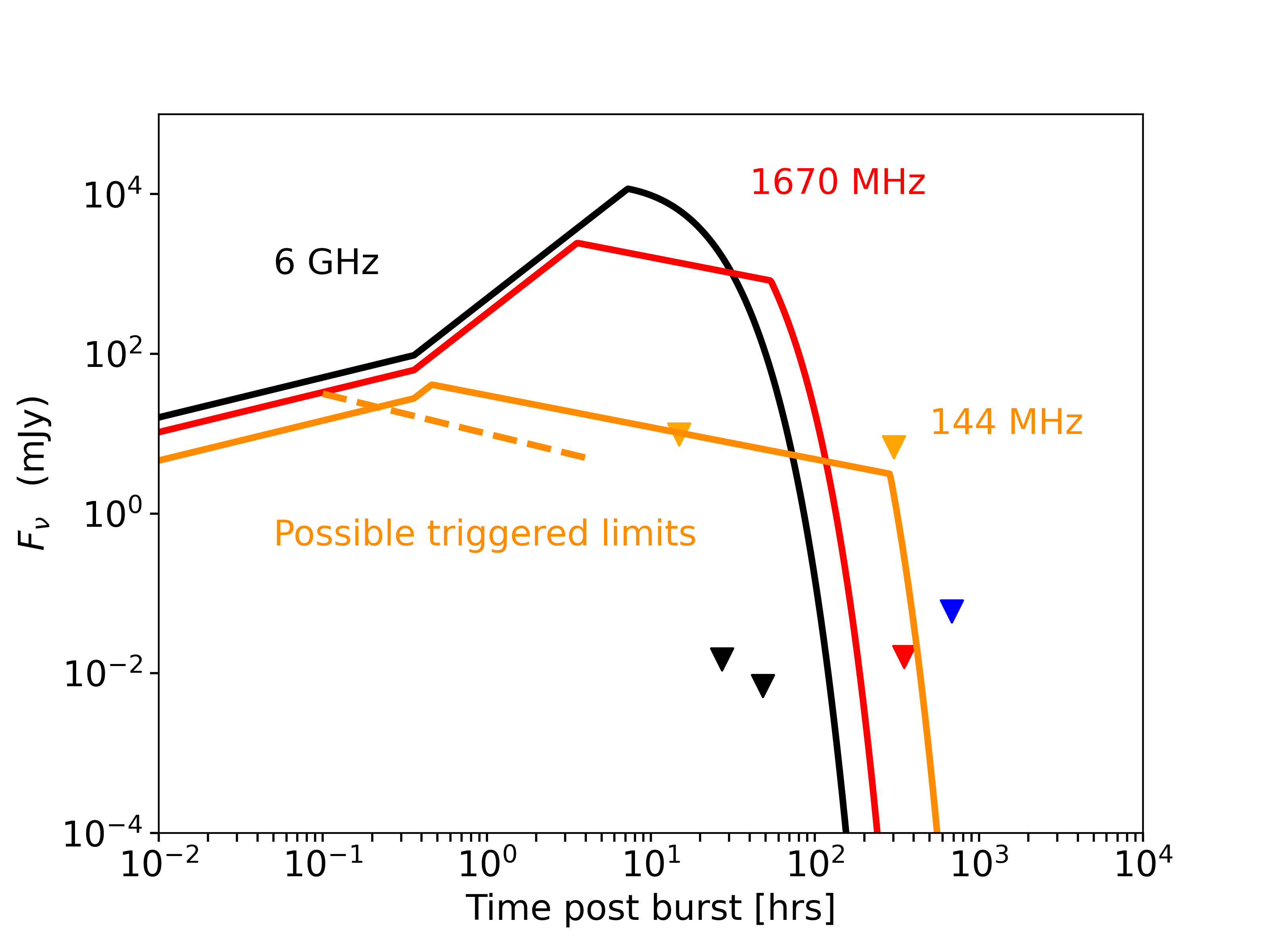}
\caption{Radio afterglow predictions for a typical luminosity FRB from SGR 1935+2154, at frequencies for which early upper limits were obtained for FRB 200428. We include the expected sensitivity from triggered observations with low-frequency radio telescopes that have rapid response capabilities, such as LOFAR or the Murchison Wide Array (MWA; \protect\citealt{2013PASA...30....7T}). We assume a conservative $5$ mJy sensitivity for a 4 hr integration at $144$ MHz, and that sensitivity increases as $\sqrt{T}$ where T is the observation time (dashed orange line). The observations are presumed to begin 6 minutes after the burst, longer than LOFAR's current rapid response capability. Inverted triangles correspond to radio upper limits at different frequencies from \protect\cite{bailes2021} as in Fig \ref{fig:sgr_1935_afterglow_lightcurve+limits}.}
\label{fig:bright_galactic_burst}
\end{figure}

In Fig. \ref{fig:bright_galactic_burst}, we present $6$ GHz, $1.67$ GHz and $144$ MHz afterglow lightcurves for a `typical' luminosity ($E_{\rm flare} = 10^{43}$ erg) FRB from within our Galaxy, along with obtained radio limits for the SGR 1934+2154. The afterglow in this Figure is calculated using the reference values of $r_{\rm FRB}$, $\Gamma_{\rm FRB}$, $t_{\rm FRB}$ \& $\nu_{\rm sync, dec}$ for a typical FRB described in Eqs. 30-32, 57 in \cite{Metzger2019}. We scale the early X-ray afterglow fluence appropriately such that the $\mathcal{F}_{\nu, \rm sync, max} \approx E_{\rm flare}/(4 \pi D^2)$. We find that of the radio limits attained after the 2020 burst, only the earliest LOFAR \& VLA limits can constrain the model in its current form. We note that for this brighter burst, the condition that $\gamma_{\rm therm} > 1$ as required by the afterglow model holds until 1000 hrs post-burst, i.e. the entire time of interest for the radio afterglow.
\par
We also show in Fig. \ref{fig:bright_galactic_burst} plausible flux limits that could have been attained if radio telescopes capable of rapid automated response such as LOFAR \citep{vanHaarlem2013,rowlinson2022} and MWA \citep{Tremblay2015,Hancock2019,Anderson2021} were triggered on the initial burst and were on source within 10 minutes. We have assumed a $5$ mJy flux limit after $4$ hr observation, where the flux limit scales as $F_{\rm lim} \propto (T_{\rm obs})^{-1/2}$ as generally expected, unless the flux threshold is limited by poor u-v coverage for very short integration times. If we are fortunate enough to observe a bonafide FRB from within our Galaxy, rapid optical and radio observations on minute to day timescales will be crucial to observe the afterglow and verify or falsify the maser shock model of FRBs. Such observations are only possible if the FRB event is reported in a timely manner on networks that distribute astronomical alerts rapidly (such as GCN \citealt{1998Barthelmy} or VOEvent \citealt{Williams2006} networks), and radio telescopes have programs in which observations can be interrupted for rapid or automatic repointing.

\subsection{Non-thermal radio afterglows}
\label{sect:shock}
Shock acceleration of particles resulting in non-thermal distributions and radiation occurs almost ubiquitously in high energy astrophysical transients such as supernovae remnants \citep{Yuan2011,Ackermann2013}, gamma-ray bursts \citep{Waxman1997}, active galactic nuclei \citep{Blandford1979} and X-ray binary jets \citep{Markoff2001}. However, particle-in-cell (PIC) simulations of relativistic, magnetized shocks suggest that particle acceleration in shocks does not occur very efficiently \citep{Sironi2009,Sironi2015}. As relativistic shocks sweep up organised magnetic field lines, they are compressed into the downstream shocked medium. As compression occurs, the angle between the magnetic field lines and the shock velocity $\theta$ increases, such that they are quasi-perpendicular $\theta \approx 90 \,\deg$. PIC simulations show no significant self-generated turbulence or magnetic field from particles in relativistic magnetized shocks \citep{Sironi2011}, and as such the particles are forced to slide along the background magnetic field lines. As these field lines run perpendicular to the shock velocity, particles do not escape the shock and therefore do not undergo repeated shock crossings required for Fermi-like shock acceleration. This is quantified in \cite{Sironi2009}, where the authors define a critical angle $\theta_{\rm crit}$, such that if $\theta > \theta_{\rm crit}$, particles would have to propagate at greater than the speed of light in order to outrun the shock and undergo multiple shock crossings as required for non-thermal acceleration. For this reason, the FRB afterglow model presented in \cite{Metzger2019,Margalit2020} assumes purely thermal radiation, nor are non-thermal particles included in Figs. \ref{fig:sgr_1935_afterglow_lightcurve+limits}, \ref{fig:afterglow_optical}, \ref{fig:afterglow_optical_windlike} or \ref{fig:bright_galactic_burst}. 
\par
As the FRB-initiating shock wave progresses it decelerates to smaller bulk Lorentz factors and propagates into regions of lower magnetic field strength (Eqs. \ref{eq:gamma_r_dec} \& \ref{eq:magnetization}). The typical magnetization and Lorentz factor values assumed in simulations of relativistic, magnetized shocks are $\sigma > 0.1$ and $\Gamma > 10$. For the SGR 1935+2154 burst, the afterglow model in \cite{Margalit2020} suggests the shock enters a non-relativistic Sedov-Taylor phase at $t_{\rm non-rel} \approx 45$ seconds via Eq. \ref{eq:t_nonrel}. This is exceptionally early due to the low-luminosity of this FRB: $t_{\rm non-rel}$ will be larger by a factor of approximately $100$ for a bright FRB, using the shock values from \cite{Margalit2020a}, corresponding to a time $t_{\rm non-rel} \approx 1.5$ hours. We suggest that in this non-relativistic phase, particles are able to undergo multiple shock crossing as has been observed in magnetized shocks of supernovae remnants. 

% In this non-relativistic phase, we suggest it is likely that particles can undergo shock acceleration in a similar manner to known magnetized, non-relativistic shocks observed in supernovae remnants. 

% In either case low-frequency radio emission extends beyond this timescale (i.e. Figs. \ref{fig:sgr_1935_afterglow_lightcurve+limits} \& \ref{fig:bright_galactic_burst}). In this phase, it is likely that particles can undergo shock acceleration in a simple manner to known magnetized, non-relativistic shocks observed in supernovae remnants. 
\par
To model the non-thermal radiation in the Sedov-Taylor phase, we assume 10\% of the total (constant) energy of the shock $E_{\rm sh} = \dfrac{4}{3} \pi R_{\rm sh}^3 n_{\rm ext} m_{\rm p} v_{\rm sh}^2 \propto t^{0}$ is available for non-thermal particle acceleration, as is canonically expected for cosmic ray acceleration in SNRs (e.g. \citealt{Strong2007}). We further assume an equipartition in energy between hadronic and leptonic acceleration. For this calculation we assume a uniform density medium throughout such that $n_{\rm ext} = n_{\rm FRB}$, noting that if the constant density does not extend to such radii, the non-thermal radiation would be lower than predicted here. Non-thermal radiation strongly depends on the magnetic field at the shock. However, in the non-relativistic regime the shock's magnetic field strength is uncertain due to the unknown magnetization of the magnetosphere-ISM transition medium, and possible amplification of the compressed magnetic field due to non-resonant Bell instability \citep{Vink2003,Bell2004}. We parameterize the magnetic field in terms of a fraction of the thermal energy density (i.e. Eq. \ref{eq:magnetization}) and the shock-amplified ISM value, such that:
\begin{equation}
    B = {\rm max}\bigg( \sqrt{64 \pi \sigma \beta^2 m_p c^2 n_{\rm ext}}\:,\: \: \chi_{\rm sh} B_{\rm ISM} \bigg)
\end{equation}
Where $\chi_{\rm sh} \approx 4$ is the shock compression ratio, the magnetization $\sigma = 0.1-1$ \citep{Metzger2019}, $\beta$ is the velocity of the blast-wave in units of the speed of light, and we assume $B_{\rm ISM} \approx 3 \; \mu$G. The minimum electron energy can be expressed in terms of the thermal particle energy:
\begin{equation}
    E_{\rm min} = \epsilon_{\rm e} \frac{(p - 2)}{2(p - 1)} \frac{m_{\rm p} \beta^2 c^2}{m_{\rm e}} + m_e c^2
\end{equation}
Where $\epsilon_{\rm e} = 0.05$ is the fraction of shock energy (i.e. half of 10\% of the total energy due to equipartition) that goes into accelerating electrons and $p$ is the slope of the power law distribution of shock-acceleration electrons which we assume to be $p = 2.2$ \citep{Sironi2013a}. The spectral luminosity in the slow cooling regime relevant for the radio afterglow is therefore:
\begin{equation}
L_{\nu} \propto \frac{K}{p+1} \nu^{-(p-1)/2} B^{(p+1)/2} 
\end{equation}
where $K = E_{\rm sh} \epsilon_{\rm e} (p-2) E_{\rm min}^{p-2}$ is the normalization factor of the non-thermal electron distribution. We note that exponential suppression of non-thermal particle radiation is assumed for times $t < t_{\rm non-rel}$, since we do not expect shock acceleration at these times (see above). We treat synchrotron self-absorption using Eq. \ref{eq:absorption}, replacing $\gamma_{\rm therm}$ with the Lorentz factor for which the critical synchrotron frequency is the observing frequency $\nu$.
\par
\begin{figure}
\centering
\includegraphics[width=0.45\textwidth]{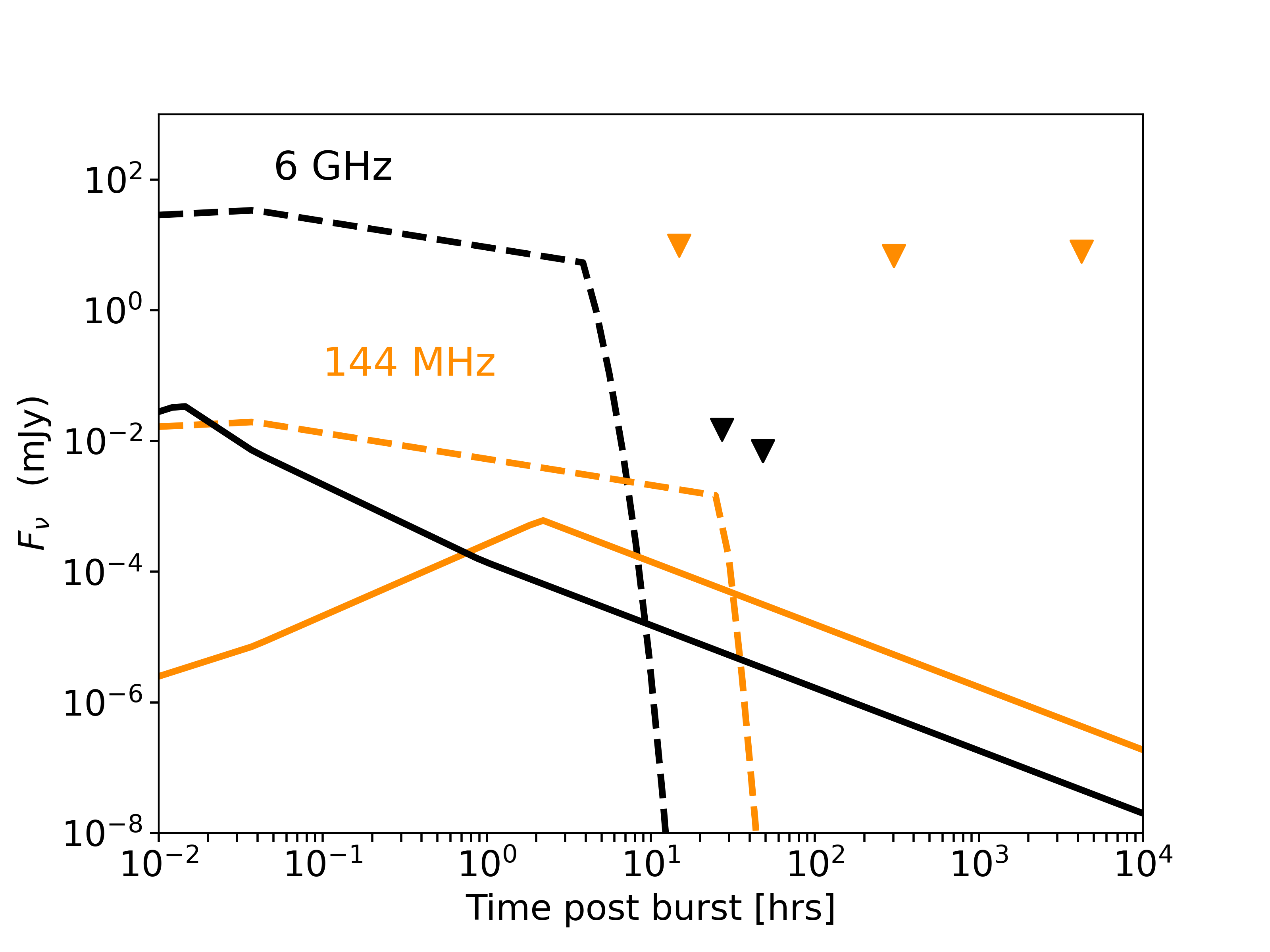}
\caption{Dotted and solid lines denote the thermal and non-thermal components respectively to the afterglow associated with the FRB from SGR 1935+2154, as described in the text. The major break denotes the time at which the emission becomes optically thin. The non-thermal afterglow contributes most significantly for lower frequencies and at later times, near the thermal synchrotron cut-off. For all times of interest, the shock magnetic field strength is always greater than the amplified ISM value. Black and orange inverted triangles correspond radio upper limits taken after FRB 200428 by VLA and LOFAR at 6 GHz and 144 MHz respectively.}
\label{fig:sgr_1935_afterglow_lightcurve+powerlaw}
\end{figure}
\begin{figure}
\centering
\includegraphics[width=0.45\textwidth]{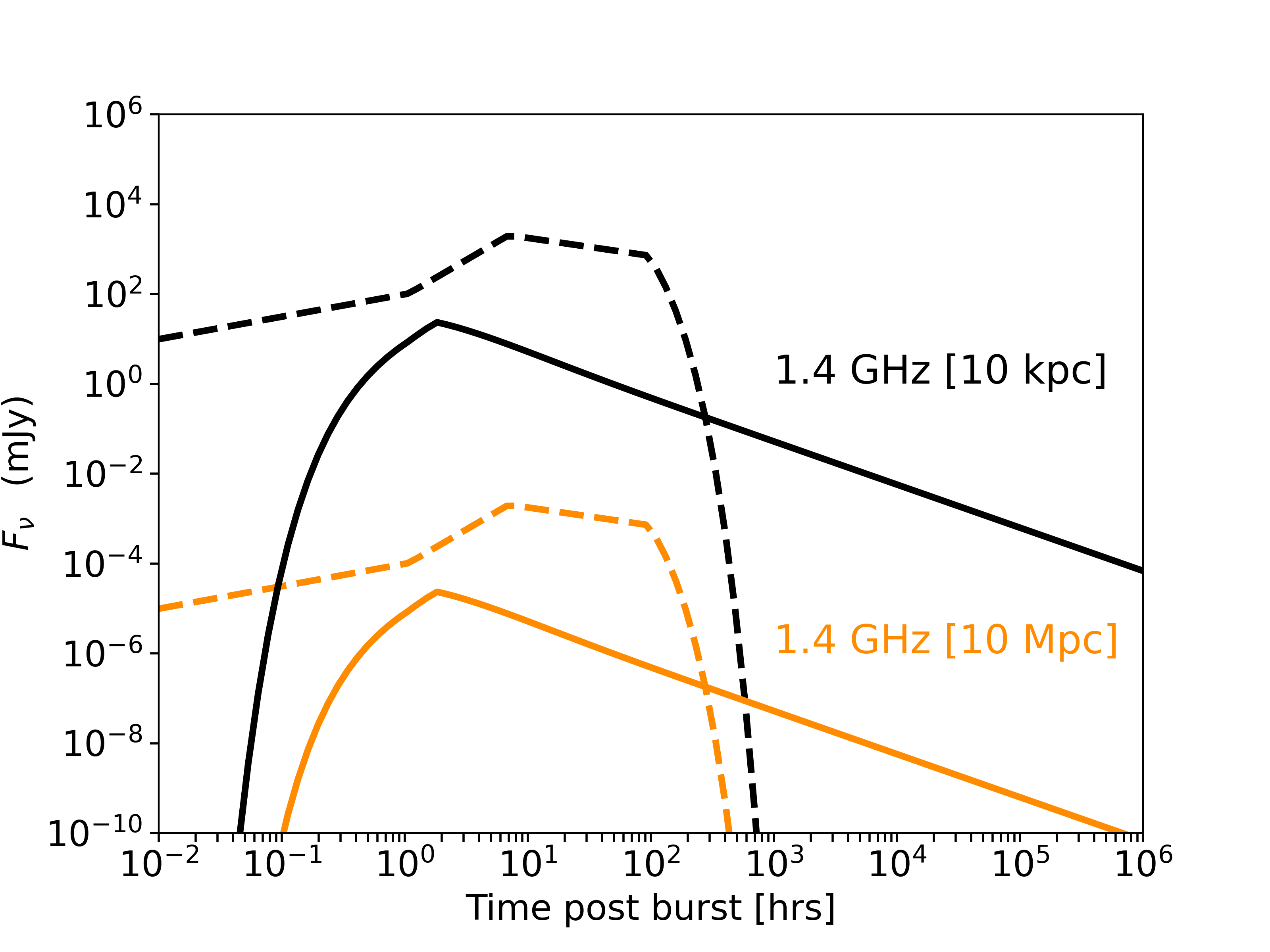}
\caption{Dotted and solid lines denote the thermal and non-thermal components respectively to the afterglow after a typical luminosity FRB, where $\nu_{\rm obs} = 1.4 \, {\rm GHz}$. We show two distances representative of a bright Galactic burst (black) and a nearby extragalactic FRB. The non-thermal emission is relatively weak, but allows for late-times monitoring of the lightcurve in the Galactic FRB case. The thermal radio afterglow is tentatively observable ($> \mu$Jy) in the hours following a close extragalactic FRB. We assume exponential suppression of non-thermal acceleration before the shock becomes non-relativistic.}
\label{fig:bright_burst_afterglow_lightcurve+powerlaw}
\end{figure}

In Fig. \ref{fig:sgr_1935_afterglow_lightcurve+powerlaw}, we show the predicted thermal (dashed) and non-thermal (solid) afterglow lightcurves for two observing frequencies. For FRB 200428, the weak burst means the afterglow is relatively dim, and does not challenge the radio upper limits previously discussed. However, we note that particularly at low frequencies the non-thermal radiation could contribute to the overall flux shortly before the thermal synchrotron cut-off. In Fig. \ref{fig:bright_burst_afterglow_lightcurve+powerlaw}, we show a similar plot for a bright FRB for two distances, corresponding to Galactic or nearby extragalactic FRBs. For a bright Galactic FRB, non-thermal radio emission could be detectable up to years after the initial burst. Long-term monitoring of such a source would provide opportunities for detailed modelling of the afterglow, contributing to our understanding of shocks more generally. However, non-thermal radio emission from even a nearby extragalactic FRB is not likely to be bright enough to probe.
\par
The closest repeating FRB resides at a distance of $D \approx 3.6$ Mpc in M81 \citep{Bhardwaj2021,Kirsten2022}. \cite{Kirsten2022} obtained deep persistent flux constraints ($6.5 \mu$Jy at 1.5 GHz) of the source before their reported FRBs, but 1-6 months after previously reported bursts by \cite{Bhardwaj2021}. Unfortunately, the low luminosity of the observed FRBs mean that predicted afterglow emission is not observable.

% \textbf{Furthermore, after $t_{\rm non-rel}$ the shock has propagated a distance approximately $d \approx c t \approx 10^{14} \, t_{\rm hour} \, {\rm cm}$. For a magnetar with a surface magnetic field of $B_s = 10^{15}$ G, at $d = 10^{14}$ cm the $B \approx 1 nG$ assuming $B \propto d^{-3}$ as we expect from dipole magnetic fields.  Therefore, at late times 1-100 hours after the shock, where radio observations may be begin. one might expect the shock to be not only non-relativistic, but also weakly magnetized.}
% Magnetic fields of $\approx$ mG have been associated with non-thermal particle acceleration in young supernova remnants:
% \cite{Vink2003} estimate the magnetic field strength at the Cassiopeia A shock front to be $B = 0.08 - 0.16$ mG, with shock accelerated electrons reaching energies of $\approx 40$ TeV. 
% \par

% Compare to SGR 1906 flare from granot

% \textbf{Possibly include lightcurves where a small number of electrons are accelerated in the shock (grb model w/Deniz and Vincent) to see whether this can be ruled out within the context of this model, given current radio limits. I could also just adhoc add a powerlaw ontop of this.}

% \begin{itemize}
%     \item modelling - should we expect persistent flux, associated source in deepest image, detect flares? \citep{Levin2020}
%     \item \citep{Yu2021} - confronting afterglow model also \citep{Yamasaki2022}
% \end{itemize}

% \subsection{SNR G57.2+0.8}
% \begin{itemize}
%     \item \textbf{This analysis will likely not make it into the thesis version}
%     \item Would LOFAR observations of the surrounding remant be interesting for distance estimates (Ping Zhou/Laura Driessen)
%     \item spectral index mapping of SNR?  \citep{Domcek2021}
%     \item Multi-wavelength observations \citep{Zhuo2020,Kothes2018}
% \end{itemize}

\section{Conclusions}
\label{sect:conclusion}
In this work we have presented results from LOFAR imaging observations of SGR 1935+2154, following the Galactic magnetar burst FRB 200428. We discuss interpretations of the LOFAR results, stressing the importance of simultaneous X-ray/radio observations of active magnetars to constrain both magnetospheric and flare/shock models of FRBs. We also make recommendations of rapid observations on minute-day timescales following Galactic or nearby extragalactic FRBs. We have analysed optical and radio early time upper limits in the context of models of FRB 200428 that predict a multi-wavelength afterglow, namely the synchrotron maser shock model. We have found that early BOOTES-3 optical observations appear to rule out simple versions of this model, but invoking a wind-like environment close to the FRB emission zone may mitigate the constraints. We also suggest that shock accelerated particle populations should be considered at late times when the shock is non-relativistic, however we find that such non-thermal emission is too faint to be observable, except for future Galactic FRBs.

\section*{Acknowledgements}
AC would like to thank Ben Margalit for a thorough explanation of the shock-maser afterglow model for the specific case of SGR 1935+2154, and for describing the calculation underlying the afterglow light curve predictions in the non-relativistic regime in their work. We would also like to thank Jeremy Harwood, Oliver Boersma \& Rhaana Starling for useful discussions and suggestions, and the anonymous referee whose comments improved this work.  
\par
Sections of this paper are based on data obtained with the International LOFAR Telescope (ILT) under project codes DDT13\_004, DDT13\_006, DDT14\_007. LOFAR is the Low Frequency Array designed and constructed by ASTRON. It has observing, data processing, and data storage facilities in several countries, that are owned by various parties (each with their own funding sources), and that are collectively operated by the ILT foundation under a joint scientific policy. The ILT resources have benefited from the following recent major funding sources: CNRS-INSU, Observatoire de Paris and Universit\'e d’Orl\'eans, France; BMBF, MIWF-NRW, MPG, Germany; Science Foundation Ireland (SFI), Department of Business, Enterprise and Innovation (DBEI), Ireland; NWO, The Netherlands; The Science and Technology Facilities Council, UK; Ministry of Science and Higher Education, Poland. K.G. acknowledges Australian Research Council (ARC) grant DP200102243.

%%%%%%%%%%%%%%%%%%%%%%%%%%%%%%%%%%%%%%%%%%%%%%%%%%
\section*{Data Availability}
The LOFAR visibility data are publicly available at the LOFAR Long Term Archive (https://lta.lofar.eu/) under the relevant project codes. A basic reproduction package providing the scripts and data required to reproduce the figures and tables of this paper will be available upon publication.



\bibliographystyle{mnras}
\bibliography{SGR1935} % if your bibtex file is called example.bib

%\begin{thebibliography}{99}
%\end{thebibliography}

%%%%%%%%%%%%%%%%%%%%%%%%%%%%%%%%%%%%%%%%%%%%%%%%%%

%%%%%%%%%%%%%%%%% APPENDICES %%%%%%%%%%%%%%%%%%%%%

% \appendix

% \section{Some extra material}

% If you want to present additional material which would interrupt the flow of the main paper,
% it can be placed in an Appendix which appears after the list of references.

%%%%%%%%%%%%%%%%%%%%%%%%%%%%%%%%%%%%%%%%%%%%%%%%%%

% Don't change these lines
\bsp	% typesetting comment
\label{lastpage}
\end{document}